\documentclass[a4paper,10pt]{article}

\title{Free and constrained symplectic integrators for numerical general relativity}

\author{Ronny Richter$^1$ and Christian Lubich$^2$\\~\\
Mathematisches Institut, Universit\"at T\"ubingen,\\
Auf der Morgenstelle 10, 72076 T\"ubingen, Germany\\
$^1$richter@na.uni-tuebingen.de, $^2$lubich@na.uni-tuebingen.de}

\usepackage{amsmath}
\usepackage{amsfonts}
\usepackage{amssymb}
\usepackage{graphicx}
\usepackage{color}
\usepackage{pstricks}

\newcommand\bfzero{{\mathbf 0}}
\newcommand\bfone{{\mathbf 1}}

\newcommand\bfc{{\mathbf c}}

\newcommand\bff{{\mathbf f}}
\newcommand\bfg{{\mathbf g}}
\newcommand\bfk{{\mathbf k}}
\newcommand\bfp{{\mathbf p}}
\newcommand\bfq{{\mathbf q}}

\newcommand\bfx{{\mathbf x}}
\newcommand\bfy{{\mathbf y}}

\newcommand\bfA{{\mathbf S}}
\newcommand\Amat{S}
\newcommand\bfD{{\mathbf D}}

\newcommand\bfI{{\mathbf I}}
\newcommand\bfalpha{{\boldsymbol \alpha}}
\newcommand\bfbeta{{\boldsymbol \beta}}
\newcommand\bfgamma{{\boldsymbol \gamma}}
\newcommand\bflambda{{\boldsymbol \lambda}}
\newcommand\bfmu{{\boldsymbol \mu}}

\newcommand\bfDelta{{\mathbf A}}
\newcommand\bfh{{\mathbf h}}
\newcommand\bfpi{{\boldsymbol\pi}}

\newcommand\bfnabla{{\boldsymbol \nabla}}
\newcommand\calH{{\cal H}}

\newcommand\dt{\Delta t}

\begin{document}


\maketitle

\begin{abstract}
We consider symplectic time integrators in
numerical General Relativity and discuss both 
free and constrained evolution schemes. For free evolution of ADM-like equations we propose the use of the St\"ormer--Verlet method, a standard symplectic integrator which here is explicit in the computationally expensive curvature terms. For the constrained evolution we give a formulation of the evolution equations that enforces the momentum constraints in a holonomically constrained Hamiltonian system and turns the Hamilton constraint function from a weak to a strong invariant of the system. This formulation permits the use of the constraint-preserving symplectic RATTLE integrator, a constrained version of the St\"ormer--Verlet method. 

The behavior of the methods is illustrated on two effectively
1+1-dimensio\-nal versions of Einstein's equations, that allow to investigate a perturbed Minkowski problem and the Schwarz\-schild space-time. We compare symplectic and non-symplectic integrators for free evolution, showing very different numerical behavior for nearly-conserved quantities in the perturbed Minkow\-ski problem. Further we compare free and constrained evolution, demonstrating in our examples that enforcing the momentum constraints can turn an unstable free evolution into a stable constrained evolution. This is demonstrated in the stabilization of a perturbed Minkowski problem with Dirac gauge, and in the suppression of the propagation of boundary instabilities into the interior of the domain in Schwarzschild space-time.
\end{abstract}


\section{Introduction}
The Einstein equations of General Relativity have a Hamiltonian formulation that arises as a consequence of the Hilbert action principle in a 3+1 slicing \cite{ADM_collection2191,Dirac-1958,Dirac-1959,Regge_Teitelboim,ashtekar}.
The present article deals with numerical methods that respect the Hamiltonian structure in the discretization.

In various areas of scientific computing, such as the dynamics of particle accelerators, molecular dynamics, celestial mechanics, quantum dynamics and electrodynamics, {\it symplectic integrators} for Hamiltonian systems have been essential for attaining favorable propagation properties (see, e.g., \cite{HaLW06,LeR04,SSC94} and references). There have also been a few papers on symplectic integrators for free evolution in general relativity
\cite{Berger_Garfinkle_CQG,Berger_Moncrief_PhysRevD.48.4676,
blanco_costa_1995,brown-2006-73,gambini-2005}. Very recently, in
\cite{Fr08} the difficulties in devising constraint-preserving symplectic integrators have been addressed but not resolved.

In this paper we present symplectic integrators for free and constrained evolution in numerical relativity and illustrate and compare their numerical properties in numerical experiments with effectively 1+1-dimensional versions of Einstein's equations.

In Section 2 we describe the general framework of the Hamiltonian formulation and its spatial discretization. For the free, ADM-like evolution we propose the {\it St\"ormer--Verlet method} in Section 3. This standard symplectic integrator (see
\cite{HaLW03,HaLW06,LeR04}) here is an implicit method, but it is explicit in the computationally expensive terms containing the discretized Ricci scalar. Moreover, the implicitness is point-wise in space and is readily resolved by simple fixed-point iteration. However, in this free evolution scheme the Hamiltonian and momentum constraints are not considered and may drift off.

In Section 4 we give a formulation of the (spatially discretized) Einstein equations that enforces the momentum constraints in a holonomically constrained Hamiltonian system. This formulation interprets the shift vector as additional momentum variables and at the same time fixes it by gauge conditions. In the spatially continuous problem, a result in
\cite{Anderson_York_4556} implies that with the enforced momentum constraints,  the Hamiltonian constraint function satisfies a conservation law. Even if this property does not extend to the space discretization, it is an extra bonus for this momentum-constrained formulation.

The problem of constraint growth in numerical relativity has also been tackled, using various techniques, in other free and constrained evolution schemes (see e.g. \cite{pretorius-2006-23,lindblom-2006-23,brodbeck-1999-40,Bonazzola_PhysRevD.70.104007,bardeen-piran-1983}). Stable evolutions of the black hole binary problem are possible since 2005 \cite{pretorius-2005-95,campanelli-2006-96,baker-2006-96,Laguna-2006,bruegmann-2006}. In the approach presented here we avoid the growth of the momentum constraints and additionally respect the Hamiltonian structure of the equations.

Although no symplectic integrators are known for general constraints in Hamiltonian systems, such integrators do exist for holonomically constrained systems. The most basic of these methods is the {\it RATTLE method}, a constrained version of the
St\"ormer--Verlet method. In Section 5 we discuss its application to the holonomically constrained formulation of Section 4.

Section 6 presents the 1+1-dimensional examples with which we have done our numerical experiments: a perturbed Minkowski problem and Schwarzschild space-time. We describe the spatial discretization, the Dirac gauge condition, and the boundary treatment that we used in our numerical experiments.

Section 7 compares a non-symplectic Runge-Kutta method and the symplectic St\"ormer--Verlet method on a perturbed Minkowski problem, showing very different dynamical behavior in the harmonic energies corresponding to different modes. These energies are almost-conserved quantities in the time-continuous problem and in the symplectic method, but completely fail to be preserved in the non-symplectic method.

Section 8 compares the free and the constrained symplectic schemes on a different perturbed Minkowski problem, illustrating that the unstable free evolution can be stabilized by enforcing the momentum constraints.

Section 9 compares free and constrained symplectic integrators on the Schwarz\-schild space-time. It turns out that in the constrained scheme, boundary instabilities do not propagate into the interior of the domain, as opposed to the free scheme.

Our numerical experiments thus illustrate remarkable properties of symplectic vs.~non-symplectic and constrained vs.~free integrators that apparently have not been addressed in the literature before.

\section{Hamiltonian formulation and space discretization}

\subsection{Super-Hamiltonian and constraints}
For classical general relativity a variety of Hamiltonian formulations is known (see \cite{franke-2006-148} for a summary of the most popular ones). Our work is based on the famous 
\emph{super-Hamiltonian} \cite{Dirac-1958,Dirac-1959} (see also \cite{Anderson_York_4556,gourgoulhon-2007}). It was discovered as a preliminary to the quantization of gravity and relies on a 3+1 splitting of space-time. The geometry is described by the 3-metric of the spatial slices and their extrinsic curvature.

In the super-Hamiltonian, the position variables are provided by the 3-metric $h_{ij}$ and the corresponding canonical momenta are denoted $\pi^{ij}$. They are related to the extrinsic curvature $K_{ij}$ by
$ \pi^{ij} = \sqrt{h}\left(K^l{}_l h^{ij} - K^{ij} \right)$.
Here $h$ is the determinant of the metric $h_{ij}$.
The super-Hamiltonian takes the form
\begin{equation}\label{super-H}
{\mathcal H} = \int (\alpha C + \beta^i C_i) \,d^3x 
\end{equation}
with freely specifiable functions $\alpha$ and $\beta^i$.
The {\it densitized lapse} $\alpha$ is related to the lapse function $N$ by $\alpha = N/\sqrt{h}$, and 
the vector with components 
$\beta^i$ is the {\it shift vector}. 
The functions $C$ and $C_i$ are given by
\begin{equation}
\label{ham-constr}
C = \pi^{ij}\pi_{ij}-\frac{1}{2}\pi^i{}_i\pi^j{}_j- h R\,,
\end{equation}
where $R$ is the Ricci scalar of the metric $h_{ij}$, and%
\footnote{Notice that $\pi^{jk}$ is a tensor density of weight $+1$.}
\begin{equation}
\label{mom-constr}
C_i = - 2 h_{ij} D_k\pi^{jk}.
\end{equation}

Solutions to the canonical Hamiltonian equations of motion
\begin{equation}
\label{eq:ADM_equations}
\dot h_{ij} = \frac{\delta{\mathcal H}}{\delta\pi^{ij}}\,,
\qquad 
\dot \pi^{ij} = - \frac{\delta{\mathcal H}}{\delta h_{ij}}
\end{equation}
are solutions of Einstein's equations if they satisfy the
{\it scalar constraint} (or Hamiltonian constraint) and the
{\it vector constraints} (or momentum constraints)
\begin{equation}\label{constraints}
C=0\, , \quad\ C_i =0\,.
\end{equation}
If these constraints are satisfied for the initial data, then they remain satisfied in the course of the evolution. However, these functions do not satisfy a conservation law for general initial data. In this sense, $C$ and $C_i$ are {\it weak invariants} but not strong invariants. In Section \ref{sect:constrained} we will consider a formulation where the momentum constraints $C_i$ are enforced (treating $\beta^i$ as dynamical variables) and the
Hamilton constraint function $C$ becomes a strong invariant.

The system \eqref{eq:ADM_equations} is the ADM system presented in \cite{ADM_collection2191}. It is weakly hyperbolic, but not strongly hyperbolic, and hence it has an ill-posed initial value problem. For practical applications one will therefore prefer to adopt another Hamiltonian formulation of general relativity, like e.g. a generalized harmonic system \cite{brown-2008}. However, we focus on the description of applications of symplectic integrators. Adopting other formulations will not change our concept, but the details will be more complicated.

For a reasonable comparison of the properties of the different time integration methods in the free evolution schemes we will consider an example where the unstable modes of the ill-posed system are not excited in the course of the simulation (see section \ref{sect:exp1}). Concerning the constrained evolution scheme (see sections \ref{sect:rattle}) we consider the modified system \eqref{constr-eom}, where the equations \eqref{holo} as well as the vector constraints are explicitly imposed. Unfortunately nothing is known yet about the well-posedness of the initial value problem of this constrained system.

\subsection{Discrete Hamiltonian -- general form}
\label{sec:Discretization_general}

To apply numerical methods one will always approximate the continuous functions $h_{ij}$, $\pi^{ij}$, $\beta^i$ and $\alpha$ through objects with finitely many degrees of freedom, and introduce discrete derivative operators that act on these finite dimensional spaces. With those ingredients a discrete Hamiltonian is derived by replacing the continuous functions and derivative operators in \eqref{super-H} with the discrete ones. A concrete example of such a finite-difference discretization is given in Section~\ref{sec:Discretisation11}. With some care in defining the discrete canonical momenta, also finite element or spectral discretizations yield finite-dimensional canonical Hamiltonian equations of motion.

We collect the discrete degrees of freedom corresponding to the functions $h_{ij}$, $\pi^{ij}$, $\alpha$, $\beta^i$ in vectors 
$\bfq$, $\bfp$, $\bfalpha$, $\bfbeta$, respectively.
The ordering in these vectors is chosen such that components corresponding to the same spatial grid point are ordered consecutively.

Since the super-Hamiltonian consists of terms that are either quadratic in the momenta $\pi^{ij}$ or linear in both $\pi^{ij}$ and $\beta^i$ or independent of the momenta, any reasonable discretization of the super-Hamiltonian $\calH$ will assume the form (we ignore the dependence on the discrete densitized lapse $\bfalpha$ in the notation)
\begin{align}
 \label{H-disc}
 H (\bfq,\bfp) = \frac12 \bfp^T \bfA(\bfq) \bfp + 
U(\bfq) + \bfbeta^T \bfD(\bfq) \bfp,
\end{align}
where $\bfA(\bfq)$ and $\bfD(\bfq)$ are matrices of the appropriate dimensions. The matrix $\bfA(\bfq)$ is a square and symmetric matrix. In a finite-difference discretization, this matrix is block-diagonal with six-dimensional blocks corresponding to the six components of $\pi^{ij}$ at a grid point, since the term in the super-Hamiltonian that is quadratic in the momenta does not contain spatial derivatives.
We then have $\bfA(\bfq)=$ blockdiag$\,(\Amat(q_\ell))$, where $q_\ell$ contains the six components of the 3-metric at the grid point $x_\ell$.

The discrete Hamiltonian is thus quadratic in the momenta, but the functional dependence on the position variables is more complicated. The computationally expensive term with the discretized Ricci scalar is subsumed in the potential $U(\bfq)$.

The canonical equations of motion for this discrete Hamiltonian then read
\begin{equation}
\label{eq:discrete_evolution_equations}
\begin{array}{rcl}
 \dot \bfq &=&  \bfA(\bfq)\bfp + \bfD(\bfq)^T\bfbeta\\[2mm]
 \dot \bfp &=& 
 -\frac12 \bfp^T \bfnabla_\bfq \bfA(\bfq)\bfp
 -\bfnabla_\bfq U(\bfq)
 -\bfbeta^T\bfnabla_\bfq \bfD(\bfq)\bfp\,.
\end{array}
\end{equation}
Without further ado, the discretized momentum constraints 
$C_i=0$, which read
\begin{equation}\label{mom-constr-disc}
\bfD(\bfq)\bfp = \bfzero\,,
\end{equation}
are {\it not} preserved under the evolution of the discretized system (\ref{eq:discrete_evolution_equations}), nor are the discretized Hamilton constraints $C=0$ preserved. There may be an exponential or even super-exponential drift away from these constraints along solutions of the discrete system 
(\ref{eq:discrete_evolution_equations}).

\section{Free evolution by the St\"ormer--Verlet method}
\label{sect:sv}

A standard symplectic integrator for Hamiltonian systems is the 
St\"ormer--Verlet scheme, which is particularly prominent in the area of molecular dynamics and enjoys a number of remarkable properties; see, e.g., \cite{HaLW03}. When applied to
(\ref{eq:discrete_evolution_equations}), a step from old values $(\bfq^n,\bfp^n)$ at time $t^n$ to values 
$(\bfq^{n+1},\bfp^{n+1})$ at time $t^{n+1}=t^n+\dt$ reads as follows:
\begin{eqnarray}
\label{sv-p1}
\bfp^{n+1/2} &=& \bfp^n - \frac\dt 2
\Big(
 \frac12 (\bfp^{n+1/2})^T \bfnabla_\bfq \bfA(\bfq^{n})\bfp^{n+1/2}
 +\bfnabla_\bfq U(\bfq^{n})
\\
\nonumber
&&  \qquad\qquad\qquad + \:
 \bfbeta^T\bfnabla_\bfq \bfD(\bfq^{n})\bfp^{n+1/2}
\Big)
\\
\label{sv-q}
\bfq^{n+1} &=& \bfq^n + \frac\dt 2 
\Big(
\bigl(\bfA(\bfq^{n+1}) + \bfA(\bfq^n)\bigr)\bfp^{n+1/2}
\\
\nonumber
&&  \qquad\qquad\quad +\: \bigl(\bfD(\bfq^{n+1})^T + \bfD(\bfq^n)^T\bigr)\bfbeta
\Big)
\\
\label{sv-p2}
\bfp^{n+1} &=& \bfp^{n+1/2} - \frac\dt 2
\Big(
 \frac12 
(\bfp^{n+1/2})^T \bfnabla_\bfq \bfA(\bfq^{n+1})\bfp^{n+1/2}
 +\bfnabla_\bfq U(\bfq^{n+1})\quad\ 
\\
\nonumber
&&  \qquad\qquad\qquad + \:
\bfbeta^T\bfnabla_\bfq \bfD(\bfq^{n+1})\bfp^{n+1/2}
\Big)\,.
\end{eqnarray}
There is only one evaluation per step of the potential $U$ that contains the computationally expensive gradient of the discretized Ricci scalar. The substeps (\ref{sv-p1}) and (\ref{sv-q}) are implicit in $\bfp^{n+1/2}$ and 
$\bfq^{n+1}$, respectively. They are solved by fixed-point iteration, which is local at every grid point.  The choice 
$\bfbeta=0$ for the shift vector further simplifies the formulas. The last substep (\ref{sv-p2}) is explicit.
We will present numerical experiments 
with this method in Sections
\ref{sect:exp1}--\ref{sect:exp3}.

The St\"ormer--Verlet integrator is a second-order method. We remark that higher-order symplectic methods are obtained by suitable compositions of steps with different step sizes; see, e.g., \cite[Section V.3]{HaLW06}.

\section{A holonomically constrained Hamiltonian formulation}
\label{sect:constrained}
Since the constraints are not taken care of in the free evolution (\ref{eq:discrete_evolution_equations}), there may be an uncontrollable drift in the discretized momentum and Hamiltonian constraints. While there exist symplectic integrators for holonomically constrained Hamiltonian systems, see \cite[Section VII.1]{HaLW06} and \cite[Chapter 7]{LeR04}, there exist no general symplectic integrators for systems where the constraints depend on both position and momentum variables, as is the case with the momentum and Hamilton constraints in general relativity. We therefore look for a reformulation of the equations of motion for general relativity that enforces the momentum constraints via holonomic constraints and, as an extra benefit, turns the Hamilton constraint function from a weak invariant into a strong invariant (i.e., satisfying a conservation law). While this reformulation can equally be done on the continuous level, we here present it for the discrete equations of motion (\ref{eq:discrete_evolution_equations}) to which we
will apply a symplectic integrator in the next section.

We now consider $\bfbeta$ as a dynamical variable and deal with it in two seemingly contradictory ways:
\begin{enumerate}
\item  We fix $\bfbeta$ by a {\it gauge condition}
\begin{equation}\label{gauge}
\bfg(\bfq)=\bfzero \quad\ \hbox{with invertible matrix }\ 
\bfDelta(\bfq) := \bfnabla_\bfq \bfg(\bfq)^T \bfD(\bfq)^T\,.
\end{equation}
Time differentiation of $\bfg(\bfq)=\bfzero$ and using \eqref{eq:discrete_evolution_equations} for $\dot\bfq$ gives
\begin{align}
\bfnabla_\bfq \bfg(\bfq)^T \bfA(\bfq)\bfp + \bfDelta(\bfq)\bfbeta = \bfzero\,,
\end{align}
which shows that indeed $\bfbeta$ is determined by the gauge condition, when $\bfDelta(\bfq)$ is invertible.

A candidate for the choice of the gauge function $\bfg$ is a discretization of the {\it Dirac gauge},
$\partial_j(h^{1/3}h^{ij})=0$. With this choice, 
$\bfDelta(\bfq)$ is a discretized second-order elliptic differential operator.
\item We consider $\bfbeta$ as {\it new momentum variables} canonically conjugate to position variables $\bfgamma$ that are not present in the Hamiltonian (\ref{H-disc}), viz.,
$$
H=H(\bfq,\bfgamma;\bfp,\bfbeta) =
\frac12 \bfp^T \bfA(\bfq) \bfp  
+ \bfbeta^T \bfD(\bfq) \bfp + 
U(\bfq)\,,
$$
which still is quadratic in the momenta $(\bfp,\bfbeta)$.
\end{enumerate}
In the free evolution, the equations of motion for this extended Hamiltonian are (\ref{eq:discrete_evolution_equations}) together with $\dot\bfbeta = -\bfnabla_\bfgamma H = 0$ and
$\dot\bfgamma = \bfnabla_\bfbeta H = \bfD(\bfq)\bfp$.

In order to enforce the discrete momentum constraints 
$\bfD(\bfq)\bfp=0$, we therefore impose the {\it holonomic constraints} (depending only on the position variables 
$(\bfq,\bfgamma)$)
\begin{equation}\label{holo}
\bfg(\bfq) = \bfzero\,,\quad\ \bfgamma=\bfzero\,.
\end{equation}
The corresponding {\it hidden constraints} obtained by time differentiation of these constraints 
and using the unchanged expressions for 
$\dot\bfq=\bfA(\bfq)\bfp + \bfD(\bfq)^T\bfbeta$ 
(see (\ref{eq:discrete_evolution_equations})) and
$\dot\bfgamma = \bfD(\bfq)\bfp$ then yields
\begin{equation}\label{hidden}
\bfnabla_\bfq \bfg(\bfq)^T \bfA(\bfq)\bfp + 
\bfDelta(\bfq)\bfbeta = \bfzero\,,\quad\
\bfD(\bfq)\bfp=\bfzero\,.
\end{equation}
This way, the discrete momentum constraints appear as hidden constraints of a holonomically constrained Hamiltonian system.
Moreover, $\bfbeta$ is completely determined by the first equation in (\ref{hidden}). 

The full set of equations of motion with Lagrange multipliers 
$\bflambda$ corresponding to the holonomic constraints
is now
\begin{equation}\label{constr-eom}
\begin{array}{rcl}
\dot\bfq &=& \bfA(\bfq)\bfp + \bfD(\bfq)^T\bfbeta
\\[2mm]
\dot\bfp &=& -\frac12 \bfp^T \bfnabla_\bfq \bfA(\bfq)\bfp
 -\bfnabla_\bfq U(\bfq)
 -\bfbeta^T\bfnabla_\bfq \bfD(\bfq)\bfp\,
- \bfnabla_\bfq \bfg(\bfq) \bflambda
\end{array}
\end{equation}
together with the constraints (\ref{holo}) and (\ref{hidden})
and formally also the equations 
\begin{equation}\label{trivial-eom}
\dot\bfgamma=\bfD(\bfq)\bfp \,,\quad\ \dot\bfbeta=-\bfmu
\end{equation} 
with Lagrange multipliers $\bfmu$ corresponding 
to the constraints $\bfgamma=0$.
It is to this formulation that we will apply a suitable numerical integrator in the next section.

This formulation, which can be similarly given also for the spatially continuous problem, does not enforce the Hamilton constraint. However, in the continuous case, a result by Anderson \& York \cite{Anderson_York_4556} shows that satisfying the momentum constraints $C_i=0$ implies that the Hamilton constraint function satisfies a conservation law
$$
(\partial_t  - {D}_\beta) C = 0\,,
$$
where $D_\beta$ is the covariant derivative operator in the direction of the shift vector $\beta=(\beta^i)$.

\section{Constrained evolution by the RATTLE method}
\label{sect:rattle}
The RATTLE method (\cite{hc_andersen}, \cite[Section VII.1]{HaLW06},
\cite[Chapter 7]{LeR04}) is an extension of the St\"ormer--Verlet method to holonomically constrained systems. It is symplectic and time-reversible, of second order accuracy, and enforces both the holonomic and the derived hidden constraints in the numerical solution. When applied to (\ref{holo})--(\ref{trivial-eom}), a step of the RATTLE method consists of the following equations, which form a nonlinear system for $\bfq^{n+1}$ and
$\bfp^{n+1}$.
\begin{enumerate}
\item First half-step for the momentum variables:
\begin{eqnarray}
\label{p-half}
\bfp^{n+1/2} &=& \bfp^n - \frac\dt 2
\Big(
 \frac12 (\bfp^{n+1/2})^T \bfnabla_\bfq \bfA(\bfq^{n})\bfp^{n+1/2}
 + \bfnabla_\bfq U(\bfq^{n})
\\
\nonumber
&&  \qquad\qquad + \:
 (\bfbeta^{n+1/2})^T\bfnabla_\bfq \bfD(\bfq^{n})\bfp^{n+1/2}
+ \bfnabla_\bfq \bfg(\bfq^{n})\bflambda^{n,+}
\Big)
\\
\label{beta-half}
\bfbeta^{n+1/2} &=& \bfbeta^n - \frac\dt 2 \bfmu^{n,+}
\end{eqnarray}
\item Full step for the position variables:
\begin{eqnarray}
\label{q-new}
\bfq^{n+1} &=& \bfq^n + \frac\dt 2 
\Big(
\bigl(\bfA(\bfq^{n+1}) + \bfA(\bfq^n)\bigr)\bfp^{n+1/2}
\\
\nonumber
&&  \qquad\qquad\quad +\: \bigl(\bfD(\bfq^{n+1})^T + \bfD(\bfq^n)^T\bigr) \bfbeta^{n+1/2}
\Big)
\\
\label{gamma-new}
\bfgamma^{n+1} &=& \bfgamma^n + \frac\dt 2 
\bigl( \bfD(\bfq^n) + \bfD(\bfq^{n+1}) \bigr)\bfp^{n+1/2}
\qquad (\bfgamma^n=\bfzero)
\end{eqnarray}
\item Position constraints:
\begin{eqnarray}
\label{q-constr}
\bfg(\bfq^{n+1}) &=& \bfzero
\\
\label{gamma-constr}
\bfgamma^{n+1} &=& \bfzero
\end{eqnarray}
\item Second half-step for the momentum variables:
\begin{eqnarray}
\label{p-half-2}
\bfp^{n+1} &=& \bfp^{n+1/2} - \frac\dt 2
\Big(
 \frac12 
(\bfp^{n+1/2})^T \bfnabla_\bfq \bfA(\bfq^{n+1})\bfp^{n+1/2}
 +\bfnabla_\bfq U(\bfq^{n+1})\quad\quad \phantom{.}
\\
\nonumber
&&  \qquad+ \:
 (\bfbeta^{n+1/2})^T\bfnabla_\bfq \bfD(\bfq^{n+1})\bfp^{n+1/2}
+ \bfnabla_\bfq \bfg(\bfq^{n+1})\bflambda^{n+1,-}
\Big)
\\
\label{beta-half-2}
\bfbeta^{n+1} &=& \bfbeta^{n+1/2} - \frac\dt 2 \bfmu^{n+1,-}
\end{eqnarray}
\item Momentum constraints:
\begin{eqnarray}
\label{beta-constr}
\bfnabla_\bfq \bfg(\bfq^{n+1})^T 
\bfA(\bfq^{n+1})\bfp^{n+1} + 
\bfDelta(\bfq^{n+1})\bfbeta^{n+1} &=& \bfzero
\\[2mm]
\label{p-constr}
\bfD(\bfq^{n+1})\bfp^{n+1}&=&\bfzero\,.
\end{eqnarray}
\end{enumerate}
Equations (\ref{p-half})--(\ref{gamma-constr}) determine 
$\bfq^{n+1}$, and (\ref{p-half-2})--(\ref{p-constr}) determine $\bfp^{n+1}$. The equations can be solved by an iterative procedure that requires only the solution of
linear systems with the matrices $\bfDelta(\bfq^n)$ and $\bfDelta(\bfq^{n+1})$ and their transposes.

\medskip\noindent
{\bf Iterative solution of (\ref{p-half})--(\ref{gamma-constr}):}
We start with $\bfq^n$ as initial iterate for $\bfq^{n+1}$,
$\bfp^n$ for $\bfp^{n+1/2}$, $\bfbeta^n$ for $\bfbeta^{n+1/2}$, and $\bfzero$ for $\bflambda^{n,+}$.
With these values we first update the iterates $\bfq^{n+1}$ and $\bfp^{n+1/2}$ through \eqref{p-half} and \eqref{q-new} respectively, inserting the initial iterates at the right-hand sides of these equations.

Then, for given iterates of
$\bfq^{n+1}$, $\bfp^{n+1/2}$ and $\bfbeta^{n+1/2}$,
equations (\ref{gamma-constr}), (\ref{gamma-new}) and 
(\ref{p-half}) yield an equation for $\bflambda^{n,+}$:
\begin{eqnarray}
\label{eq:lambda_np}
&&\frac 1 2 \Bigl( \bfD(\bfq^n) + \bfD(\bfq^{n+1}) \Bigr)
\bfnabla_\bfq \bfg(\bfq^{n})\,\bflambda^{n,+}
\\ \nonumber &&\quad\ = \frac1\dt
\Bigl( \bfD(\bfq^n) + \bfD(\bfq^{n+1}) \Bigr)
\biggl(
\bfp^n - \frac\dt 2
\Big(
 \frac12 (\bfp^{n+1/2})^T \bfnabla_\bfq \bfA(\bfq^{n})\bfp^{n+1/2}
\\ \nonumber
\nonumber
&&  \qquad\qquad\qquad\qquad\qquad\qquad
+\: \bfnabla_\bfq U(\bfq^{n}) + 
 (\bfbeta^{n+1/2})^T\bfnabla_\bfq \bfD(\bfq^{n})\bfp^{n+1/2}\Big)\biggr)\,.
\end{eqnarray}
The matrix on the left-hand side is $O(\dt)$-close to
$\bfDelta(\bfq^n)^T$ and therefore invertible under condition (\ref{gauge}).

Equation \eqref{eq:lambda_np} can be interpreted as the requirement to choose $\bflambda^{n,+}$ such that the momentum constraints are satisfied for the next iterate, $\bfp^{n+1/2}_{{\mathrm{next}}}$. Instead of using \eqref{p-half} to obtain $\bfp^{n+1/2}_{{\mathrm{next}}}$ from the current iterates we may also assume that it is approximately
$\bfp^{n+1/2}_{{\mathrm{next}}}=\bfp^{n+1/2}-\dt/2\bfnabla_\bfq\bfg(\bfq^n)\Delta\bflambda^{n,+}$ with a small correction $\Delta\bflambda^{n,+}$ to the Lagrange multipliers. This leads to the following equation,
\begin{eqnarray}
\label{eq:Deltalambda_np}
&&\bfDelta(\bfq^n)^T\Delta\bflambda^{n,+} =
\frac1\dt\Bigl( \bfD(\bfq^n) + \bfD(\bfq^{n+1}) \Bigr)\bfp^{n+1/2}.
\end{eqnarray}
We solve this approximately for the increment $\Delta\bflambda^{n,+}$, and replace the current iterate
$\bflambda^{n,+}:=
\bflambda^{n,+}+\Delta\bflambda^{n,+}$.

Inserting (\ref{q-new}) into (\ref{q-constr}) gives an equation for $\bfbeta^{n+1/2}$. With the current iterates in the argument of $\bfg$, we solve for the increment
$\Delta\bfbeta^{n+1/2}$ in a step of a simplified Newton iteration:
\begin{align}
\label{eq:Deltabeta_n12}
\bfDelta(\bfq^n)\Delta&\bfbeta^{n+1/2} 
= -\frac1\dt\bfg\left(
\bfq^{n+1}\right)\\
\nonumber
& +
\frac{\dt}{4}\bfnabla_\bfq\bfg(\bfq^{n+1})^T\left(\bfA(\bfq^{n+1}) + \bfA(\bfq^n)\right)\bfnabla_\bfq\bfg(\bfq^{n})\Delta\bflambda^{n,+}
\end{align}
and replace $\bfbeta^{n+1/2}:=
\bfbeta^{n+1/2}+\Delta\bfbeta^{n+1/2}$.
We then update the iterates of $\bfp^{n+1/2}$ and $\bfq^{n+1}$ using \eqref{p-half} and (\ref{q-new}) respectively with the current iterates on the right-hand side.
We thus have the following schematic iteration cycle:
$$
\bflambda^{n,+} \longrightarrow
\bfbeta^{n+1/2} \longrightarrow
\bfp^{n+1/2},\,\bfq^{n+1}\quad\hbox{and iterate}.
$$
\medskip\noindent
{\bf Solution of (\ref{p-half-2})--(\ref{p-constr}):}
Inserting (\ref{p-half-2}) into (\ref{p-constr}) yields
a linear system for $\bflambda^{n+1,-}$ with the matrix
$\bfDelta(\bfq^{n+1})^T$. After solving this system we compute
$\bfp^{n+1}$ from (\ref{p-half-2}), and then $\bfbeta^{n+1}$
is obtained from solving the linear system
(\ref{beta-constr}) with the matrix $\bfDelta(\bfq^{n+1})$.

We remark that equations (\ref{beta-half}) and (\ref{beta-half-2}) are ignored, since they only determine approximations to the
Lagrange multipliers $\bfmu=-\dot\bfbeta$ that are without further interest.

\medskip\noindent
{\bf Symplecticity.} The RATTLE method is symplectic with
respect to the canonical symplectic two-form 
$d\bfq\wedge d\bfp+d\bfgamma\wedge d\bfbeta$ for the
{\it extended} phase space of $(\bfq,\bfgamma;\bfp,\bfbeta)$
restricted to the constraint manifold; 
see \cite[Section VII.1]{HaLW06},
\cite[Chapter 7]{LeR04}. Because of $d\bfgamma=\bfzero$, the extended symplectic two-form here actually reduces to the original canonical symplectic two-form $d\bfq\wedge d\bfp$.

\section{1+1 dimensional test cases}
\label{sect:11}

After the general discussion in the previous sections we now describe in detail the problems considered in our numerical experiments. We derive a simplified Hamiltonian in Section \ref{sec:simple_Hamiltonian}, and in Section \ref{sec:Discretisation11} we describe the spatial discretization procedure that leads to a discrete Hamiltonian. In Section \ref{sec:11gauge} we introduce the gauge conditions that we use and discuss some problems that are related to that topic. We describe a straightforward boundary treatment in Section \ref{sec:boundary_conditions}, before we turn to the particular test problems in Section \ref{sec:free_test_scenarios}.

\subsection{Simplified continuous Hamiltonian}
\label{sec:simple_Hamiltonian}

For a first test we restrict to simple problems. We therefore consider only those solutions of Einstein's equations that satisfy the following requirements:
\begin{align}
\label{eq:requirements_metric}
 \nonumber
 h_{ij}&\equiv 0\quad \mbox{for}\;i\neq j,\\
 \nonumber
 h_{ij}(x^1,x^2,x^3)&= h_{ij}(x^1,\bar x^2,\bar x^3)\quad\forall x^2,\bar x^2,x^3,\bar x^3;
 \;i,j=1,2\\
 h_{33}&\equiv \zeta h_{22},
\end{align}
where either $\zeta\equiv 1$ or $\zeta=\sin^2 x^2$. The latter case 
corresponds to spherically symmetric space-times, and  the class of solutions with $\zeta\equiv 1$ includes a perturbed Minkowski geometry. Both cases will be considered later on.

It is easy to show that one obtains solutions in either of these classes if \eqref{eq:requirements_metric} is satisfied at the initial hypersurface $\Sigma_0$, if moreover at $\Sigma_0$
\begin{align}
\label{eq:requirements_init}
 \nonumber
 \pi^{ij}&\equiv 0\quad\mbox{for}\;i\neq j,\\
 \nonumber
 \pi^{ij}(x^1,x^2,x^3)\zeta^{-1/2}&= \pi^{ij}(x^1,\bar x^2,\bar x^3)\bar\zeta^{-1/2}\quad\forall x^2,\bar x^2,x^3,\bar x^3;\;
 i,j=1,2\\
 \pi^{33}&\equiv \zeta^{-1}\pi^{22},
\end{align}
(where $\bar\zeta=1$ or $\bar\zeta=\sin^2 \bar x^2$ in the two cases, respectively)
and if the gauge is chosen such that everywhere
\begin{align}
\label{eq:requirements_gauge}
 \nonumber
 \beta^1(x^1,x^2,x^3) &= \beta^1(x^1,\bar x^2,\bar x^3)\quad\forall x^2,\bar x^2,x^3,\bar x^3,&
 \beta^2 &\equiv 0 \equiv \beta^3,\\
 \alpha(x^1,x^2,x^3)\zeta^{1/2} &= \alpha(x^1,\bar x^2,\bar x^3)\bar\zeta^{1/2}
 \quad\forall x^2,\bar x^2,x^3,\bar x^3.
\end{align}
To summarize, if \eqref{eq:requirements_gauge} is satisfied at every time slice and if
\eqref{eq:requirements_metric},\eqref{eq:requirements_init} are satisfied at $\Sigma_0$ then \eqref{eq:requirements_metric},\eqref{eq:requirements_init} are satisfied for all times, as long as Einstein's equations hold.

Because of the last equations in \eqref{eq:requirements_metric},\eqref{eq:requirements_init} it is natural to define
\begin{align}
 \tilde h &:= \frac12\left(h_{22}+\zeta^{-1} h_{33}\right),&
 \tilde \pi &:= \pi^{22}+\zeta\pi^{33}.
\end{align}
It turns out that with these definitions $\tilde\pi$ is indeed the canonical momentum corresponding to $\tilde h$.

In the spherically symmetric case we now consider the equatorial hypersurface, i.e., $x^2=\pi/2$ and $\zeta=1$.
Using the new variables it can be shown that the following simplified Hamiltonian provides the correct evolution equations for the functions $h_{11}$, $\pi^{11}$, $\tilde h$ and $\tilde \pi$:
\begin{align}
\label{eq:simple_Hamiltonian}
\nonumber
 \mathcal H&=\int dx^1\bigg[
    \alpha\left(
    \frac12\pi^{11}\pi^{11}h_{11}h_{11}-
    \pi^{11}\tilde\pi h_{11}\tilde h\right)\\
    \nonumber&\qquad\qquad
    -\alpha\left(
    \frac12 \partial_1 \tilde h\partial_1 \tilde h-
    2 \tilde h \partial_1^2 \tilde h+
    \tilde h\partial_1 \tilde h\partial_1 \log(h_{11})+2\xi h_{11}\tilde h\right)\\
    &\qquad\qquad + 2\pi^{11}h_{11}\partial_1\beta^1+
    \pi^{11}\beta^1\partial_1 h_{11}+
                \tilde\pi\beta^1\partial_1 \tilde h
    \bigg],
\end{align}
where $\xi=1$ in the spherically symmetric case and $\xi=0$ if $\zeta\equiv 1$.

This Hamiltonian has a similar structure to the super-Hamiltonian \eqref{super-H}, its discretization will therefore also be of the form \eqref{H-disc}.

\subsection{Space discretization in the 1+1 dimensional setting}
\label{sec:Discretisation11}

Based on \eqref{eq:simple_Hamiltonian} we now derive a discrete Hamiltonian. 
We introduce two uniform%
\footnote{The grids are uniform with respect to the spatial coordinate system, i.e. $x_{i+1}-x_i=\Delta x$ and $\bar x_{i+1}-\bar x_i=\Delta \bar x=\Delta x$.}
spatial grids $\{x_1,\ldots,x_N\}$ and $\{\bar x_1,\ldots,\bar x_{M}\}$,%
\footnote{If we apply periodic boundary conditions then $M=N$, otherwise $M=N-1$.}
staggered such that $\bar x_i=(x_i+x_{i+1})/2$. 
We approximate the functions $f=\alpha$, $h_{11}$, $\pi^{11}$, $\tilde h$, $\tilde \pi$ by piecewise constant functions $f^D$, such that 
$$
f^D(x) = f_i\quad\ \hbox {for }\ x_i-\Delta x/2\leq x < x_i+\Delta x/2,
$$ 
and then replace the continuous functions in $\mathcal H$ by the piecewise constant ones ($f\rightarrow f^D$). For the shift function $\beta^1$ the staggered grid $\{\bar x_1,\ldots,\bar x_{M}\}$ is the basis of this discretization, i.e.,
$$
(\beta^1)^D(x) = \beta_i\quad\ \hbox {for }\ 
\bar x_i-\Delta x/2\leq x < \bar x_i+\Delta x/2.
$$
Discretizing $\beta^1$ on the staggered grid leads to better results in the constrained evolution.

Spatial derivatives are approximated by those piecewise constant functions whose functional values are obtained through centered finite differencing, i.e.,
\begin{align}
\nonumber
 (\partial_1 f)^D(x) &= \left(f_{i+1}-f_{i-1}\right)/(2\Delta x),\\
(\partial_1^2 f)^D(x) &=
\left(f_{i+1}-2f_{i}+f_{i-1}\right)/\Delta x^2
\end{align}
for $x_i-\Delta x/2\leq x < x_i+\Delta x/2$. An analogous formula is used also for the derivatives of $\beta^1$ on the grid $\{\bar x\}$.
These functions again replace the continuous functions $\partial_1 f$ and $\partial_1^2 f$,
respectively, in the continuous Hamiltonian $\mathcal H$.

To get the discretization of the integrand in \eqref{eq:simple_Hamiltonian} one additionally needs to take a logarithm and to perform additions as well as multiplications. These are pointwise operations where the continuous functions can be replaced by the piecewise constant ones easily.
Hence, the integrand of the discrete Hamiltonian $H$ is again a piecewise constant function.

Comparing the discrete Hamiltonian for the 1+1 dimensional setting with its general form \eqref{H-disc} we identify the potential $U(\bfq)$ with the discretization of the second line in \eqref{eq:simple_Hamiltonian}.
The matrix $\bfA$ is derived from the kinetic term
\begin{align}
\label{eq:origin_of_A}
\int dx^1\alpha\left(\frac12\pi^{11}\pi^{11}h_{11}h_{11}-
    \pi^{11}\tilde\pi h_{11}\tilde h\right),
\end{align}
and the matrix $\bfD$ comes from the shift term
\begin{align}
\label{eq:origin_of_D}
\int dx^1 \left(2\pi^{11}h_{11}\partial_1\beta^1+
    \pi^{11}\beta^1\partial_1 h_{11}+
                \tilde\pi\beta^1\partial_1 \tilde h \right).
\end{align}

As we discussed in section \ref{sec:Discretization_general}, the matrix $\bfA$ is blockdiagonal with each block corresponding to a single grid point, because \eqref{eq:origin_of_A} does not contain spatial derivatives. Here, in the 1+1-dimensional setting, the blocks are two dimensional corresponding to the functions $\pi^{11}$ and $\tilde\pi$ respectively, and the elements of the blocks can be obtained through
\begin{align}
 \Amat_{\pi^{11}\pi^{11}}&=\alpha h_{11} h_{11},&
 \Amat_{\pi^{11}\tilde\pi}&=-\alpha h_{11}\tilde h,&
 \Amat_{\tilde\pi\tilde\pi}&=0.
\end{align}

Concerning the matrix $\bfD$, the integrand in \eqref{eq:origin_of_D} does contain derivatives. Since we use a staggered grid for the discretization of 
$\beta^1$, here $\bfD$ becomes an $M\times 2N$ matrix of the form
\begin{align}
\nonumber
\label{eq:form_of_D}
&\qquad\quad M=N-1&
&M=N\mbox{ (periodic boundary conditions)}\\
 &\left(
\begin{array}{cccccc}
 * & *  &&&& \\
 & * & * &&&0 \\
 && \ddots & \ddots && \\
 0&&& * & * & \\
 &&&& * & *
\end{array}
\right)&
&\qquad \quad \left(
\begin{array}{cccccc}
 * & *  &&&& \\
 & * & * &&&0 \\
 && \ddots & \ddots && \\
 &&& * & * & \\
 &0&&& * & * \\
 *&&&&& *
\end{array}
\right),
\end{align}
where each star represents a row-vector
\begin{align}
\label{eq:block_matrix_D}
 D&=\left(
 \begin{array}{cc}
  D_{\beta^1 \pi^{11}}, & D_{\beta^1 \tilde\pi}
 \end{array}
\right).
\end{align}

\subsection{Gauge conditions}
\label{sec:11gauge}

For the constrained evolution scheme discussed in section \ref{sect:constrained} we additionally need to choose a gauge constraint $\bfg(\bfq)=\bfzero$. In what follows we derive this function from the Dirac gauge \cite{Dirac-1959}, a condition that in the continuous problem fixes the spatial coordinate system. This gauge condition is formulated in terms of a flat background metric, see \cite{Bonazzola_PhysRevD.70.104007} for details.

In the 1+1 dimensional context the Dirac gauge condition becomes
\begin{align}
\label{eq:Dirac_gauge}
\partial_1\left((x^1)^{-4\xi/3}h_{11}^{-2/3}\tilde h^{2/3}\right) = 0,
\end{align}
where $\xi=1$ in the spherically symmetric case and $\xi=0$ for the $\zeta\equiv 1$ class of problems. The dependence on $x^1$ for spherically symmetric systems comes from the flat background metric, which is the Minkowski metric in spherical coordinates there.

From the continuous gauge condition \eqref{eq:Dirac_gauge} we derive a discrete one as follows. First we observe that we obtain equation \eqref{eq:Dirac_gauge} when we require that the variation of the following integral with respect to $\lambda$ vanishes
\begin{align}
\label{eq:discr_Dirac_gauge}
 \mathcal H_C:=\int dx^1\lambda\,\partial_1\left((x^1)^{-4\xi/3}h_{11}^{-2/3}\tilde h^{2/3}\right).
\end{align}

This integral is again discretized by the procedure described in Section \ref{sec:Discretisation11}, that is, the functions $h_{11}$, $\tilde h$ and $x^1$ are approximated by piecewise constant functions corresponding to the grid $\{x_1,\ldots, x_N\}$ and $\lambda$ becomes a piecewise constant function based on the staggered grid $\{\bar x_1,\ldots,\bar x_{M}\}$.
We thus obtain a discrete function $H_C$ of the grid variables $\bfq$, $\bfx$ and $\bflambda$. Following the procedure in the continuous case we take the derivative of $H_C$ with respect to $\bflambda$ as the discrete gauge constraints.

This defines functions $\bfg(\bfq)$, but as discussed in section \ref{sect:constrained} we additionally need that the matrix $\bfDelta(\bfq)$ is invertible. It is not a priori clear that this condition is satisfied. Therefore we made several spot checks by calculating the singular values of $\bfDelta(\bfq)$.

It turns out that in the simulations where the computational domain possesses boundaries, in particular for spherical symmetry, we indeed find that the ratio of the largest to the smallest singular value was at most $2\cdot 10^4$ and the ratio of the smallest to the second smallest was always below $5$. But with periodic boundary conditions it turns out that one singular value is more than ten orders of magnitude smaller than the others, and hence $\bfDelta(\bfq)$ must be regarded as singular.

\medskip\noindent
{\bf Constraints for periodic boundary conditions.}
The origin of the non-trivial kernel of $\bfDelta(\bfq)$ can be found already in the continuous 1+1 dimensional formulation. There the expression $\bfDelta(\bfq)\bfbeta$ translates to
\begin{align}
A(h)\beta^1 &:= \frac{2}{3}\partial_1\left(
\left(\frac{h_{11}}{\tilde h}\right)^{2/3}\left[\left(\frac{\partial_1\tilde h}{\tilde h} - \frac{\partial_1 h_{11}}{h_{11}}\right)\beta^1
-2\partial_1\beta^1\right]
\right).
\end{align}
It can be easily checked that with periodic boundary conditions the equations
\begin{align}
\int dx^1A(h)f&=0, &
\int dx^1 f A(h)\left(\tilde h^{1/2}h_{11}^{-1/2}\right) &= 0
\end{align}
hold for any function $f$. That means the constant functions are in the kernel of the transpose operator $A(h)^T$, and functions of the form $c\tilde h^{1/2}h_{11}^{-1/2}$ ($c=\mathrm{const.}$) are in the kernel of $A(h)$. If the Dirac gauge is satisfied then $\tilde h/h_{11}=\mathrm{const}$., so that the constant functions are also in the kernel of $A(h)$.

This argument can also be extended to other gauge conditions and probably to higher dimensions. We do not want to elaborate on this here, we only mention that for periodic boundary conditions $A(h)$ will have a non-trivial 
kernel for any pointwise gauge condition of the form $g(h_{11},\tilde h)=0$.

Now, the singularity of $\bfDelta(\bfq)$ means that we cannot solve equations \eqref{eq:Deltalambda_np}
and \eqref{eq:Deltabeta_n12} for $\Delta\bflambda$ and $\Delta\bfbeta$ in the RATTLE scheme. We therefore solve another system which enforces the discrete momentum constraints only up to a multiple of the spatially constant vector $\bfone_N=(1,\ldots,1)^T\in\mathbb R^N$. It turns out that for $\bfg(\bfq)=\bfzero$ this vector is indeed in the kernels of $\bfDelta(\bfq)$ and $\bfDelta(\bfq)^T$, and that the system
\begin{align}
\left(
\begin{array}{cc}
\bfDelta(\bfq) & \bfone_N\\
\bfone_N^T & 0
\end{array}
\right)\left(
\begin{array}{c}
\Delta\bfbeta\\
\bar\beta
\end{array}
\right)
&=
\left(
\begin{array}{c}
\bff\\
0
\end{array}
\right)
\end{align}
has a unique solution for any right-hand side $\bff$. We then solve systems of this form instead of \eqref{eq:Deltalambda_np} and \eqref{eq:Deltabeta_n12}, and take the obtained $\Delta\bflambda$ and $\Delta\bfbeta$ as the right correction to the Lagrangian multipliers and the shift variables respectively.

However, with that procedure we do not have control of the mean value of the momentum constraints. We see this immediately when we notice that
$\bfone^T_N\bfD(\bfq)(\bfp-\bfnabla_\bfq\bfg(\bfq)\bflambda)=\bfone^T_N\bfD(\bfq)\bfp-\bfone^T_N\bfDelta(\bfq)^T\bflambda$
is independent of $\bflambda$ (since $\bfone_N$ is in the kernel of $\bfDelta(\bfq)$). Hence, the mean value of the momentum constraints cannot be brought to zero by this procedure, and instead of satisfying \eqref{mom-constr-disc}, we can only fulfill $\mathbf P_{\bfone}\bfD(\bfq)\bfp=\bfzero$ 
where $\mathbf P_{\bfone}=\bfI_{N\times N}-\bfone_N\bfone_N^T/N$ is the projector to the subspace orthogonal to $\bfone_N$.

The full discrete momentum constraints $\bfD(\bfq)\bfp=\bfzero$ 
could be enforced by adding an extra gauge condition as a discretization of an integral condition
$$
\int c(h_{11},\tilde h) f\, dx^1 = 0
$$
by
$$
\bff^T \bfc(\bfq) =\bfzero,
$$
where the function $\bfc(\bfq)$ and the vector $\bff$ are chosen such that
$$
\bfone_N^T \bfD(\bfq)\bfnabla_\bfq \bfc(\bfq)\bff \ne 0.
$$
It turns out that this condition corresponds to
$$
\int c(h_{11},\tilde h) \,\partial_1 f\, dx^1 \ne 0
$$
in the continuous case.
We have, however, not included such an extension by an integral gauge condition in our numerical
experiments.

\subsection{Boundary treatment}
\label{sec:boundary_conditions}

We now describe the implementation of boundary conditions. As we focus on the investigation of the time evolution, we tried to keep this aspect simple.

For the tests in the $\zeta\equiv 1$ class of solutions we impose periodic boundary conditions and there is actually no boundary. But in the spherically symmetric case it is not possible to impose periodic boundary conditions.

For our spatial discretization, the time derivatives of the variables at the grid point $x_i$ only depend on the variables at the neighboring grid points $x_{i-2}$, $x_{i-1}$, $x_{i}$, $x_{i+1}$, $x_{i+2}$ and $\bar x_{i-2}$,
$\bar x_{i-1}$, $\bar x_{i}$, $\bar x_{i+1}$.

\begin{figure}
\begin{center}
\begin{picture}(0,0)%
\includegraphics{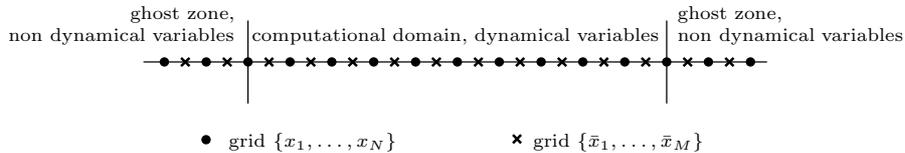}%
\end{picture}%
\setlength{\unitlength}{4144sp}%
\begingroup\makeatletter\ifx\SetFigFontNFSS\undefined%
\gdef\SetFigFontNFSS#1#2#3#4#5{%
  \reset@font\fontsize{#1}{#2pt}%
  \fontfamily{#3}\fontseries{#4}\fontshape{#5}%
  \selectfont}%
\fi\endgroup%
\begin{picture}(5159,915)(706,-3035)
\put(2026,-2986){\makebox(0,0)[lb]{\smash{{\SetFigFontNFSS{7}{8.4}{\familydefault}{\mddefault}{\updefault}{\color[rgb]{0,0,0}grid $\{x_1,\ldots,x_N\}$}%
}}}}
\put(3826,-2986){\makebox(0,0)[lb]{\smash{{\SetFigFontNFSS{7}{8.4}{\familydefault}{\mddefault}{\updefault}{\color[rgb]{0,0,0}grid $\{\bar x_1,\ldots,\bar x_M\}$}%
}}}}
\put(1441,-2221){\makebox(0,0)[lb]{\smash{{\SetFigFontNFSS{7}{8.4}{\familydefault}{\mddefault}{\updefault}{\color[rgb]{0,0,0}ghost zone,}%
}}}}
\put(4681,-2356){\makebox(0,0)[lb]{\smash{{\SetFigFontNFSS{7}{8.4}{\familydefault}{\mddefault}{\updefault}{\color[rgb]{0,0,0}non dynamical variables}%
}}}}
\put(4681,-2221){\makebox(0,0)[lb]{\smash{{\SetFigFontNFSS{7}{8.4}{\familydefault}{\mddefault}{\updefault}{\color[rgb]{0,0,0}ghost zone,}%
}}}}
\put(2161,-2356){\makebox(0,0)[lb]{\smash{{\SetFigFontNFSS{7}{8.4}{\familydefault}{\mddefault}{\updefault}{\color[rgb]{0,0,0}computational domain, dynamical variables}%
}}}}
\put(721,-2356){\makebox(0,0)[lb]{\smash{{\SetFigFontNFSS{7}{8.4}{\familydefault}{\mddefault}{\updefault}{\color[rgb]{0,0,0}non dynamical variables}%
}}}}
\end{picture}%
 \caption{The two spatial grids in the computational domain and in the ghost zones.}
 \label{fig:spatial_grid}
\end{center}
\end{figure}

Therefore we extend the spatial grids beyond the computational domain, that is, we introduce ghost zones to the left and to the right of the boundary (see Figure \ref{fig:spatial_grid}). The function values at the grid points in these ghost zones are then treated as non-dynamical variables. In general these function values are fixed through the choice of boundary conditions, and here we simply set them to values that we read off from a reference solution.

The discrete Hamiltonian still has the form \eqref{H-disc}, we only change the interpretation of the variables. Therefore the equations of motion for the variables in the computational domain have the same form as in the case of periodic boundary conditions and the variables in the ghost zones do not possess evolution equations.  

In the general form of the Hamiltonian \eqref{H-disc} we then have to distinguish between dynamical variables in the computational domain and 
non-dynamical variables in the ghost zones. In particular we must subdivide the matrix $\bfD$ such that the non-dynamical character of the ghost variables can be considered. Then, $\bfD$ takes the form (cf. equation \eqref{eq:form_of_D})
\begin{align}
\label{eq:D_decomposition}
 \bfD =
\left(
\begin{array}{ccc}
 \widehat \bfD^1 & \bar \bfD^1 & \widehat \bfD^2 \\
 \tilde \bfD^1 & \bfD_{\rm int} & \tilde \bfD^2 \\
 \widehat \bfD^3 & \bar \bfD^2 & \widehat \bfD^4
\end{array}
\right),
\end{align}
where $\bfD_{\rm int}$ corresponds to grid points in the computational domain and the remaining matrices to grid points in the ghost zones.

For $\bfA$ the consideration of ghost variables leads to an analogous result, but $\bfA$ is block diagonal and hence the only non vanishing ``ghost matrices'' are $\widehat \bfA^1$ and $\widehat \bfA^4$. These matrices only depend on variables in the ghost zones. Hence, they have no influence on the equations of motion and are therefore irrelevant. We collect the ghost matrices in three matrices
\begin{align}
\nonumber
 \tilde \bfD &:= \left(
\begin{array}{cc}
 \tilde \bfD^1 & \tilde \bfD^2
\end{array}
\right),\\
\nonumber
 \bar \bfD &:= \left(
\begin{array}{cc}
 \bar \bfD^1\\
 \bar \bfD^2
\end{array}
\right),\\
 \widehat \bfD &:= \left(
\begin{array}{cccc}
 \widehat \bfD^1 & \widehat \bfD^2\\
 \widehat \bfD^3 & \widehat \bfD^4
\end{array}
\right).
\end{align}

If we denote the variables in the ghost zones as $\tilde \bfp$, $\tilde \bfq$ and $\tilde\bfbeta$, respectively, then the Hamiltonian can be written as
\begin{align}
 \label{eq:ghost_general_form_Hamilton}
\nonumber
 H &= \frac12 \bfp^T \bfA_{\mathrm{int}}(\bfq) \bfp + U(\bfq,\tilde \bfq)\\
&\qquad
 + \bfbeta^T \bfD_{\rm int}(\bfq,\tilde \bfq) \bfp
 + \bfbeta^T \tilde \bfD(\bfq,\tilde \bfq) \tilde \bfp
 + \tilde\bfbeta^T \bar \bfD(\bfq,\tilde \bfq) \bfp
 + \tilde\bfbeta^T \widehat \bfD(\bfq,\tilde \bfq) \tilde \bfp.
\end{align}
From this Hamiltonian one obtains the equations of motion for $\bfq$ and $\bfp$, as well as the one-step map in the St\"ormer--Verlet scheme as described in Sections \ref{sec:Discretization_general} and \ref{sect:sv}, respectively.

We note that with the chosen space discretization, $\tilde\bfD(\bfq,\tilde\bfq)=0$. Moreover, it turns out that $\bfD_{\rm int}(\bfq,\tilde \bfq)$ does not depend on $\tilde \bfq$, and so the
discrete momentum constraints become
$$
\bfD_{\rm int}(\bfq)\bfp = \bfzero
$$ 
and thus do not depend on the variables in the ghost zones.
This structure is due to the discretization of the shift 
$\beta^1$ at the staggered grid $\{\bar x\}$. If the shift is discretized at $\{x\}$ then $\bfD$ does not have the bidiagonal structure \eqref{eq:form_of_D}, but there are more 
non-vanishing elements, and the momentum constraints depend 
on $\tilde\bfq$ and $\tilde\bfp$ as well.

The grid functions in the ghost zones are specified as
\begin{align}
\label{eq:Dirichlet_boundary}
 f(\tilde x,t)=f_{\mathrm ext}(\tilde x,t)
\end{align}
for each grid function $f$, each grid point in the ghost zone $\tilde x$ and all times $t$, with given reference functions $f_{\mathrm ext}$
in the exterior. This does not apply to the Lagrange multipliers $\bflambda$,
which are defined only on the staggered grid in the interior domain.

In our example of the Schwarzschild space-time given below, the analytically 
known stationary solution is chosen as initial data and exterior data, leaving
only a numerical dynamics that reflects the stability properties of the various numerical schemes.

\subsection{Test scenarios}
\label{sec:free_test_scenarios}

In the following sections we investigate the properties of the 
St\"ormer--Verlet and the RATTLE method in two different situations. The first one, a perturbed Minkowski space-time, 
is an example for the case $\zeta\equiv 1$, and the second one, the Schwarzschild space-time, is a spherically symmetric solution.

\paragraph{Perturbed Minkowski.}

The Minkowski metric describes a flat space-time, the analytical solution is (with $x=x^1$, $y=x^2$, $z=x^3$)
\begin{align}
 g=-dt^2+dx^2+dy^2+dz^2.
\end{align}
It is easy to check that for $t=\,$const. slicing this solution really is in the class \eqref{eq:requirements_metric} with $\zeta\equiv 1$ and thus the numerical schemes we described are applicable.

Here we perturb the Minkowski initial data.
We denote the perturbations $\varepsilon$ and $\delta$, and get that at a slice $\Sigma_t$ the independent components of the 3-metric are
\begin{align}
 h_{11} &= 1+\varepsilon_{11},& \tilde h&=1+\tilde\delta.
\end{align}
The canonical momenta become
\begin{align}
 \pi^{11}&=\delta^{11},&\tilde\pi&=\tilde\varepsilon,
\end{align}
and without perturbations the slicing density is identical one and the shift vector vanishes:
\begin{align}
 \alpha &\equiv 1+\varepsilon,&\beta^1 &= \varepsilon^1.
\end{align}

The spatial grid is chosen as the uniform grid
\begin{align}
 x_i = \frac{i}{N}-\frac{1}{2N}-0.5,\quad i=1,\ldots,N,
\end{align}
and we apply periodic boundary conditions.

\paragraph{Schwarzschild space-time in isotropic coordinates.}

The Schwarzschild space-time in isotropic coordinates is described by the metric (with $R=x^1$, $\theta=x^2$, $\phi=x^3$)
\begin{align}
 g=-\frac{(M-2R)^2}{(M+2R)^2}dt^2+\frac{(M+2R)^4}{16R^4}\left(dR^2+R^2d\Omega^2\right),
\end{align}
where $d\Omega^2 = d\theta^2+\sin^2\theta\, d\phi^2$. For $t=\,$const. slicing we thus obtain
\begin{align}
 h_{11} &= \frac{(M+2R)^4}{16R^4},&
 \tilde h &= \frac{(M+2R)^4}{16R^2}.
\end{align}
The extrinsic curvature and the canonical momenta vanish as well as the shift,
\begin{align}
 \pi^{11} &= 0,&
 \tilde \pi &= 0,&
 \beta^1 &= 0,
\end{align}
and the densitized lapse is
\begin{align}
\label{eq:Schwarzschild_alpha}
 \alpha &= \frac{64 R^4(2R-M)}{(2R+M)^7}.
\end{align}

The spatial grid is chosen as 
\begin{align}
 R_i=1+\frac{i-1}{N-1},\quad i=1,\ldots,N.
\end{align}
To treat the boundaries we continue the uniform spatial grids beyond the boundaries, introducing $K$ grid points in ghost zones to the left and the right.

The  Schwarzschild solution in isotropic coordinates  satisfies Dirac gauge. This permits us to obtain initial data easily and to compare the numerical results with the exact stationary Schwarzschild solution.
Applying a numerical method with these initial data
yields a purely numerical dynamics that reflects the stability properties of the various numerical schemes.

\section{A perturbed Minkowski problem: symplectic vs. non-symplectic integrators}
\label{sect:exp1}

\paragraph{Symplectic St\"ormer--Verlet vs.~non-symplectic ICN scheme.}
We compare the results of the St\"ormer--Verlet scheme (see Section \ref{sect:sv}) with that of a free non-symplectic scheme, the iterated Crank Nicholson method (ICN). For a system of ordinary differential equations $\dot \bfy = \bff(t,\bfy)$, ICN reads
\begin{align}
\nonumber
 \bfk_1 &=\frac\dt{2}\,\bff(t_n,\bfy^n),\\
\nonumber
 \bfk_2 &=\frac\dt{2}\,\bff(t_n+\Delta t/2,\bfy^n+\bfk_1),\\
\nonumber
 \bfk_3 &=\Delta t \,\bff(t_n+\Delta t/2,\bfy^n+\bfk_2),\\
 \bfy^{n+1} &= \bfy^n + \bfk_3.
\end{align}
In both schemes the densitized lapse as well as the shift are chosen not to change with time, $\dot\bfalpha=\bfzero$ and $\dot\bfbeta=\bfzero$.

We apply the schemes to the perturbed Minkowski example of Section \ref{sec:free_test_scenarios}, where the perturbation is chosen such that the functions denoted $\varepsilon$ are Gaussian functions with width $1/20$, center $0$ and height $10^{-3}$. The functions denoted $\delta$ are chosen to vanish. That is, we perturb $h_{11}$, $\tilde \pi$, $\alpha$ and $\beta^1$, but we keep $\tilde h$ and $\pi^{11}$ from the unperturbed Minkowski problem. The case where all functions are perturbed by random noise, is discussed in the next section.

\paragraph{Simulation data.}
The reason to choose $\delta=0$ here is the hyperbolicity of the considered system. It turns out that the equations of motion that correspond to \eqref{eq:simple_Hamiltonian} are weakly hyperbolic, but not strongly hyperbolic in the sense of \cite{Gundlach_2006}. However, if we have $\tilde h=1$ and $\pi^{11}=0$ at some time $t_0$ then the continuous as well as the discrete evolution equations lead to $\tilde h\equiv 1$ and $\pi^{11}\equiv 0$. We can hence omit these functions in the analysis of hyperbolicity. The obtained system for $h_{11}$ and $\tilde\pi$ is then strongly hyperbolic.

In the simulations we use grids with $51$ or $201$ grid points. The size of the time step is the same as the spatial grid spacing, i.e., $\Delta t = \Delta x$. 
The computational costs are comparable for both time-stepping methods.

\paragraph{Simulation results.}
It turns out that for both integrators the simulations are stable at least until $t=1000$. We do not see any growing modes, and we expect that the simulations remain stable for a much longer time. 
The discrete momentum constraint vanishes identically, 
because $\pi^{11}$ as well as $\partial_1 \tilde h$ vanish. The discrete Hamilton constraint does not vanish exactly, but it stays almost constant at about $5\cdot 10^{-13}$ for both schemes.

To see how the perturbations evolve, we calculate the \emph{harmonic energies} corresponding to $h_{11}$. Let $\bfh_{11}$ be the vector composed of the grid values of $h_{11}$, and let $\widehat \bfh_{11}^k$ denote the $k$th component of its discrete Fourier transform. Let further $\widehat{\dot \bfh}{}_{11}^k$ be the $k$th component of the Fourier transform of $\dot \bfh_{11}=\bfnabla_{\bfpi^{11}}H$. Then, the harmonic energy in the $k$th mode is $E_k = \frac12(|\widehat{\dot \bfh}{}_{11}^k|^2 + k^2|\widehat{\bfh}{}_{11}^k|^2)$.

\begin{figure}
\begin{center}
 \begin{tabular}{cc}
  \includegraphics[scale=0.65]{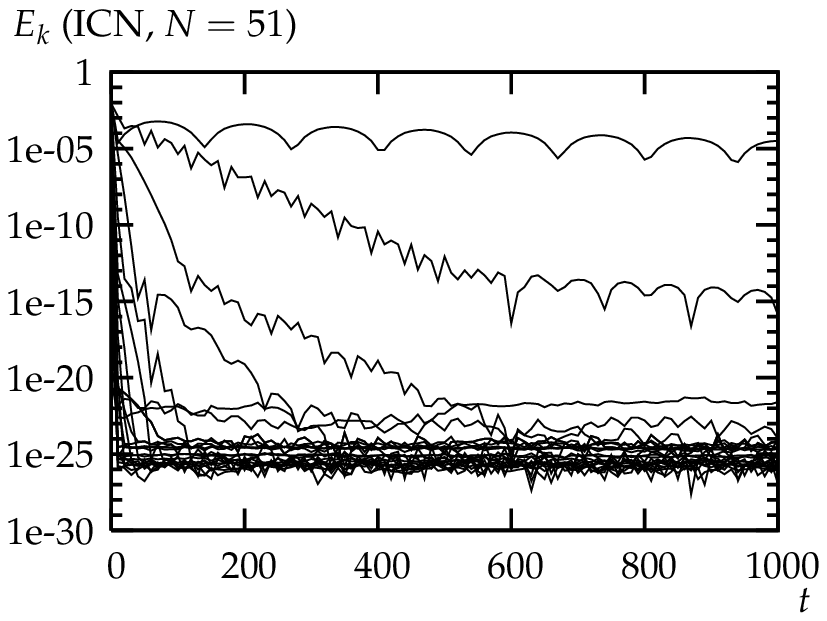}
  \includegraphics[scale=0.65]{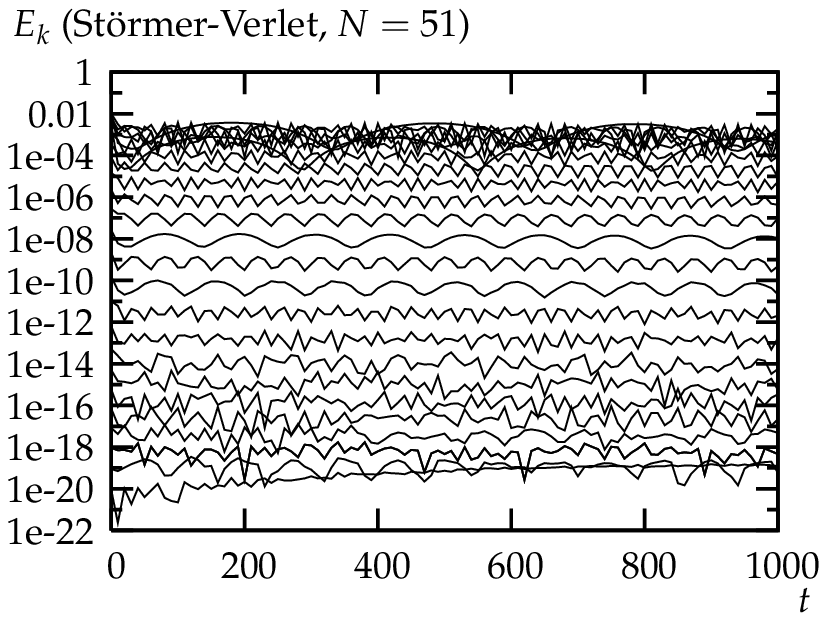}&
 \end{tabular}
 \caption{Harmonic energies in the simulations with 51 grid points. Left: Results of the iterated Crank Nicholson scheme, Right: results of the St\"ormer--Verlet scheme.}
 \label{fig:exp1:harmonic_energies_ICN_SV}
\end{center}
\end{figure}

As we see in Figure \ref{fig:exp1:harmonic_energies_ICN_SV}, the $E_k$ are approximately preserved by the St\"ormer--Verlet scheme over long times, whereas with the ICN scheme they incorrectly decay exponentially until they reach the level of 
round-off error. The exponential decay is much faster for high frequencies.

Very similar results are obtained in the computations with 201 grid points. It turns out that for even numbers of grid points the highest frequency of the system grows quadratically for symplectic as well as non-symplectic evolution, whereas the remaining Fourier modes show the same behavior as in the case of odd numbers of grid points.  

\paragraph{Discussion of results.}
To sum up the essential feature in this numerical experiment, 
we have seen that the harmonic energies are preserved for long times by the St\"ormer--Verlet scheme, whereas they decay quickly in the ICN scheme. 
The symplectic St\"ormer--Verlet scheme thus reproduces the propagation of perturbations in this problem remarkably better than the ICN scheme or in fact any explicit Runge-Kutta scheme.

\section{A perturbed Minkowski problem: stabilization by constraints}
\label{sect:exp2}

In this section we compare results of the St\"ormer--Verlet scheme of Section \ref{sect:sv} and the RATTLE scheme of Section \ref{sect:rattle} with constraints imposed as in Section~\ref{sec:11gauge}. In both schemes we choose the densitized lapse not to change with time, $\dot\bfalpha=\bfzero$, and in the St\"ormer--Verlet scheme also the shift is independent of time, $\dot\bfbeta = \bfzero$.

\paragraph{Simulation data.}
Again we apply these schemes to the perturbed Minkowski example of Section \ref{sec:free_test_scenarios} with periodic boundary conditions, but here we choose the perturbation functions 
$\varepsilon$ as well as $\delta$ to be noise in the interval $(-2.5\cdot 10^{-7}/N^2,2.5\cdot 10^{-7}/N^2)$. In particular, 
$\delta\neq 0$ so that the free evolution equations are only weakly hyperbolic. We choose the time step 
$\Delta t = \Delta x$, and apply the schemes for $N=50$ as well as $N=200$ grid points.
This example is the robust stability test suggested in \cite{Alcubierre:2003pc}.

\begin{figure}
\begin{center}
 \begin{tabular}{cc}
 \includegraphics[scale=0.65]{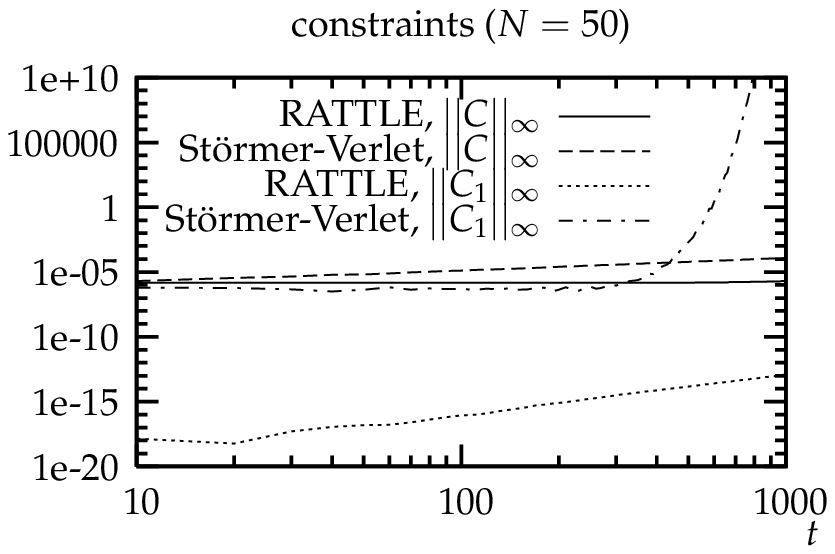}&
 \includegraphics[scale=0.65]{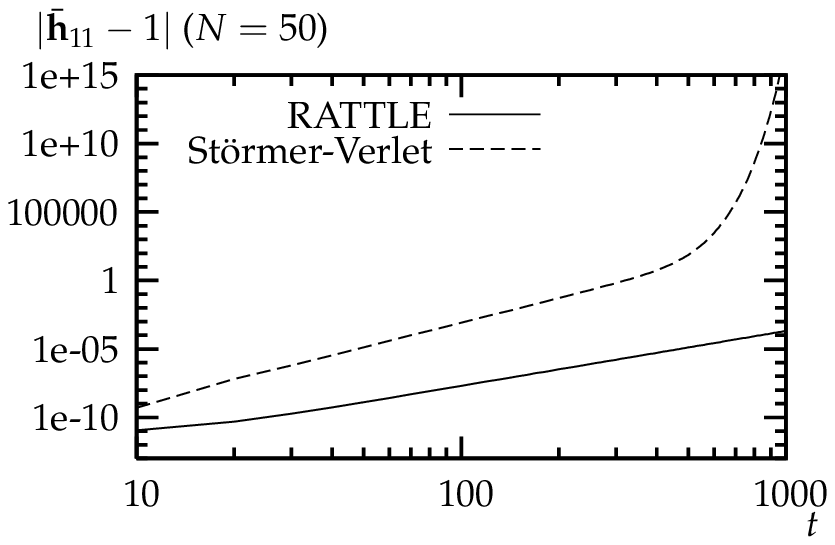}
\end{tabular}
\caption{Behavior of the numerical solution. Left: The maximum norms of the Hamilton and momentum constraints $C$ and $C_1$, right: the evolution of the mean value of $\bar{\bfh}_{11}$ of $\bfh_{11}$.}
\label{fig:exp2:constraints_mean_val_h11}
\end{center}
\end{figure}

\paragraph{Simulation results.}
We first investigate the constraints. We see in Figure \ref{fig:exp2:constraints_mean_val_h11} that the maximum norm of the Hamilton constraint function grows linearly in the St\"ormer--Verlet scheme and is almost constant at about $10^{-6}$ for the RATTLE scheme. Moreover, in the RATTLE scheme the shape of the Hamilton constraint function changes very little with time.

Concerning the momentum constraint function, in the St\"ormer--Verlet scheme its maximum norm stays constant at about $10^{-6}$ until $t\approx 200$ and grows superexponentially afterwards. The RATTLE scheme on the other hand provides very small momentum constraints of about $10^{-18}$-$10^{-14}$ that grow cubically with time.
A closer look reveals that the momentum constraint function in the RATTLE scheme is nearly constant already after the first timestep, and the cubic growth is due to a growth of the mean value (see Section~\ref{sec:11gauge}).

From Figure \ref{fig:exp2:constraints_mean_val_h11} we also see that in the St\"ormer--Verlet scheme the error of the mean value of $\bfh_{11}$ (we denote this mean value $\bar\bfh_{11}$) grows with order six in the beginning, before a superexponential growth becomes important. In the beginning the errors of the functions $\bfh_{11}$, $\tilde\bfh$ and $\tilde\bfpi$ themselves grow cubically, linearly and quadratically respectively, while the error of $\bfpi^{11}$ stays nearly constant. However, at later times the errors of $\bfh_{11}$ and $\bfpi^{11}$ grow superexponentially.

In the RATTLE scheme the error of the mean value $\bar\bfh_{11}$ grows quartically. The function $\bfh_{11}$ itself is almost constant in space, and its error is basically due to the error of the mean value. It turns out that the errors of the functions $\tilde\bfh$ and $\tilde\bfpi$ grow linearly, and the error of $\bfpi^{11}$ stays nearly constant.

\begin{figure}
\begin{center}
 \begin{tabular}{cc}
 \includegraphics[scale=0.65]{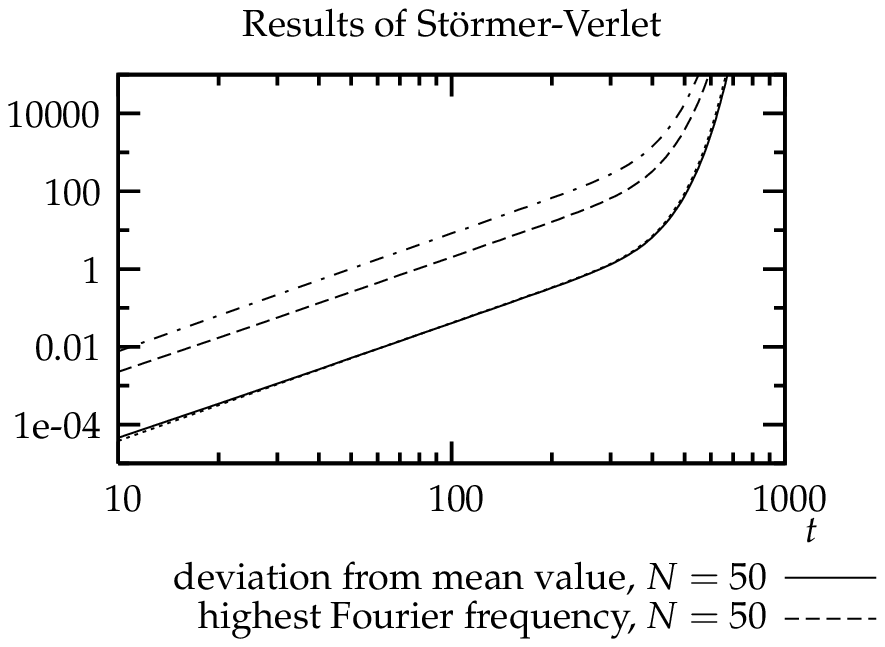}&
 \includegraphics[scale=0.65]{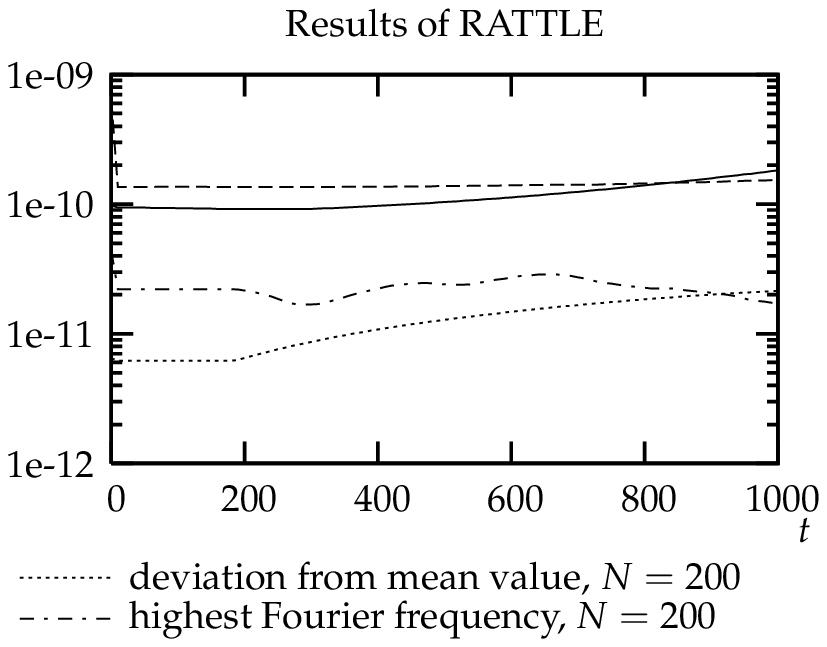}
\end{tabular}
\caption{The maximal deviation of $\mathbf h_{11}$ from its mean value, i.e. $||\mathbf h_{11}-\bar{\mathbf h}_{11}||_\infty$, and the highest frequency component of the Fourier transform $\widehat{\mathbf h}_{11}$. Left: results of the St\"ormer--Verlet scheme. Right: results of the RATTLE scheme.}
\label{fig:exp2:deviation_high_freq}
\end{center}
\end{figure}

We observe that the main errors of $\bfh_{11}$ in the RATTLE scheme come from errors of its mean value. Since it is also interesting to see how the remaining components behave, we investigate the deviations from the mean value separately. In Figure \ref{fig:exp2:deviation_high_freq} we see that the maximal deviation of $\bfh_{11}$ from its mean value grows cubically in the beginning for the St\"ormer--Verlet scheme. Comparing that deviation in the coarse grid ($N=50$) and in the fine grid ($N=200$), we do not see any difference at all. In the RATTLE scheme the deviation is at most $2\cdot 10^{-10}$ and it is more than one order of magnitude smaller in the fine grid.

We also look at high-frequency errors of $\bfh_{11}$. We see in Figure \ref{fig:exp2:deviation_high_freq} that in the St\"ormer--Verlet scheme the component of $\widehat\bfh_{11}$ that corresponds to the highest frequency grows cubically in time. Moreover it turns out that this component is about four times larger in the fine grid than in the coarse grid. In the RATTLE scheme these high frequency errors are very small, and moreover they are about four times smaller in the fine grid.

\paragraph{Discussion of results.}
We have seen that in the RATTLE schemes the error of the Hamilton constraint stays almost constant. This is in agreement with the result by Anderson and York \cite{Anderson_York_4556} that for the continuous problem the Hamilton constraint function satisfies a conservation law when the momentum constraints are satisfied (cf. also the comment at the end of Section \ref{sect:constrained}).
It is remarkable that also in the St\"ormer--Verlet scheme the Hamilton constraint function grows only linearly, although the momentum constraint is not satisfied and grows superexponentially at the end of the simulation.

In the RATTLE scheme we force the momentum constraints to be satisfied, but we discussed in section \ref{sec:11gauge} that we do not have control about their mean value. The numerical results show that even this mean value is small.
It is interesting that for the RATTLE scheme, the errors of the function $\bfh_{11}$ are  also mainly due to errors of its mean value.

In the St\"ormer--Verlet scheme the high frequency errors grow cubically with time and become larger in the fine grid. These errors probably trigger the superexponential growth that occur at the end of the simulation. It is typical for discretized weakly hyperbolic systems that high frequency errors become larger when the grid is refined (see e.g. \cite{calabrese_1228487}). High frequency errors do not play a significant role in the RATTLE scheme. They  remain small and become even smaller when more grid points are used for the spatial discretization. This example illustrates that imposing constraints may stabilize the problem.

\section{Schwarzschild space-time: free vs. constrained scheme}
\label{sect:exp3}

In this section we again compare results of the St\"ormer--Verlet and the RATTLE method. As in section \ref{sect:exp2} we choose the densitized lapse to satisfy $\dot\bfalpha=\bfzero$ in both schemes, and in the St\"ormer--Verlet scheme also $\dot\bfbeta = \bfzero$ is fulfilled.

\paragraph{Simulation data.}
We apply the schemes to the Schwarzschild space-time (see Section \ref{sec:free_test_scenarios}). Again we choose the time step $\Delta t = \Delta x$, and use grids with $N=51$ as well as $N=201$ grid points. We impose boundary conditions as described in Section \ref{sec:boundary_conditions}.

\begin{figure}
\begin{center}
 \begin{tabular}{cc}
 \includegraphics[scale=0.65]{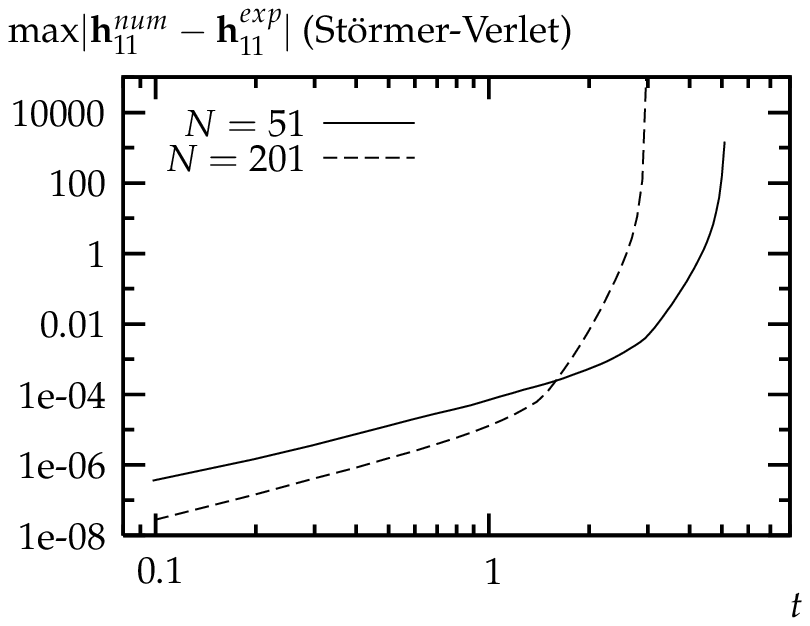}&
 \includegraphics[scale=0.65]{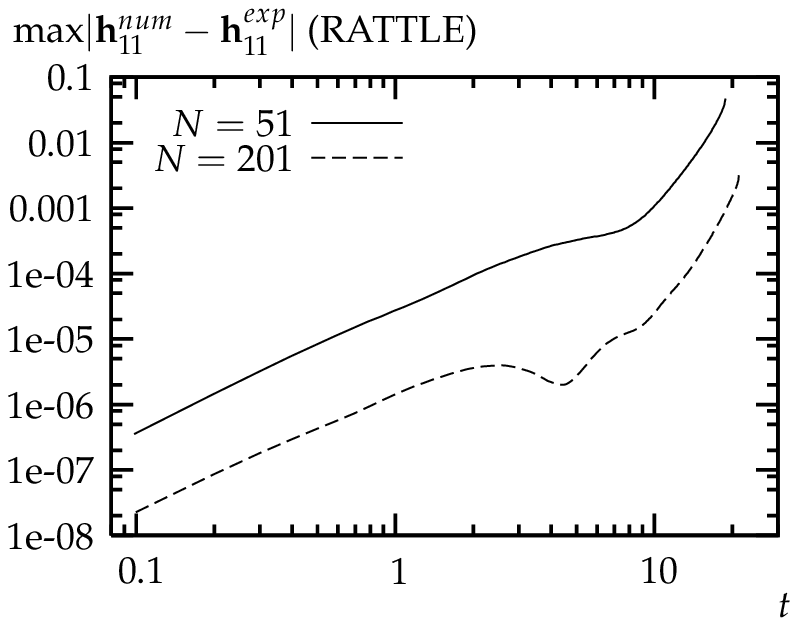}
\end{tabular}
\caption{The difference between the numerical and the analytical solution $\bfh_{11}^{num}-\bfh_{11}^{exp}$ for the Schwarzschild space-time. Left: results of the St\"ormer--Verlet scheme, right: results of the RATTLE scheme.}
\label{fig:exp3:error_hRR}
\end{center}
\end{figure}

\paragraph{Simulation results.}
From Figure \ref{fig:exp3:error_hRR} we see that the error of the numerical solution grows quadratically in the beginning but at some point before the evolution breaks down, a more rapid growth becomes important.
The period of quadratical growth of errors is for the St\"ormer--Verlet scheme in the order of $M$ (the mass of the black hole) and it is longer for the coarser grid. With the RATTLE scheme, the rapid growth of errors starts at about $10M$, and moreover the errors in the fine grid are always smaller than those in the coarse grid.

A closer look reveals that in the results of St\"ormer--Verlet there are high frequency errors near the boundary. These errors propagate from the boundary into the interior of the computational domain and further amplify. We also observed that small perturbations of the initial data as well as the use of, e.g., ICN instead of St\"ormer--Verlet have negligible effects on the solution.
In the RATTLE scheme high frequency errors of the functions $\bfh_{11}$, $\tilde \bfh$, $\bfpi^{11}$ and $\tilde\bfpi$ are very small.

\begin{figure}
\begin{center}
 \begin{tabular}{cc}
 \includegraphics[scale=0.65]{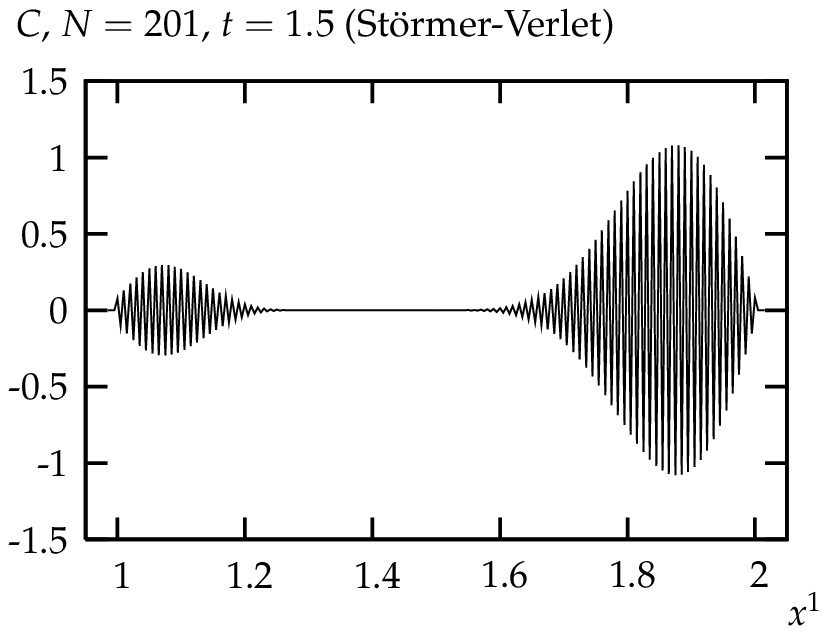}&
 \includegraphics[scale=0.65]{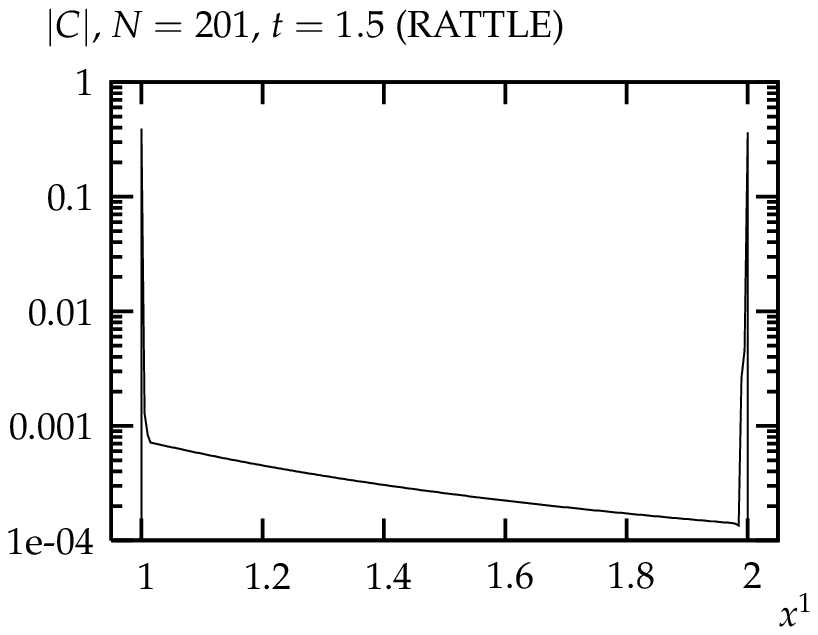}
\end{tabular}
\caption{The discrete Hamilton constraint function in the St\"ormer--Verlet scheme (left) and in the RATTLE scheme (right) at $t=1.5$ for $N=201$ grid points.}
\label{fig:Schw_Hamilton_constraint_function}
\end{center}
\end{figure}

\begin{figure}
\begin{center}
 \begin{tabular}{cc}
 \includegraphics[scale=0.65]{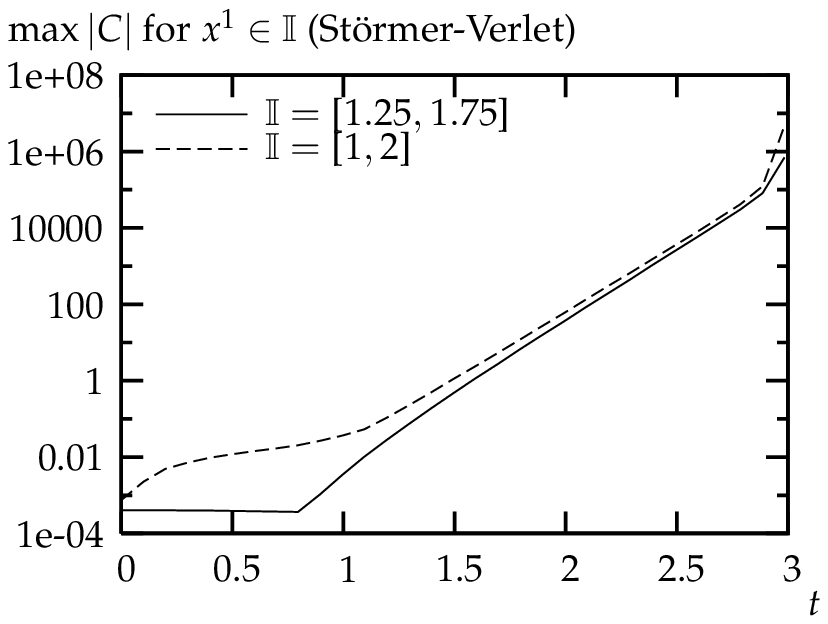}&
 \includegraphics[scale=0.65]{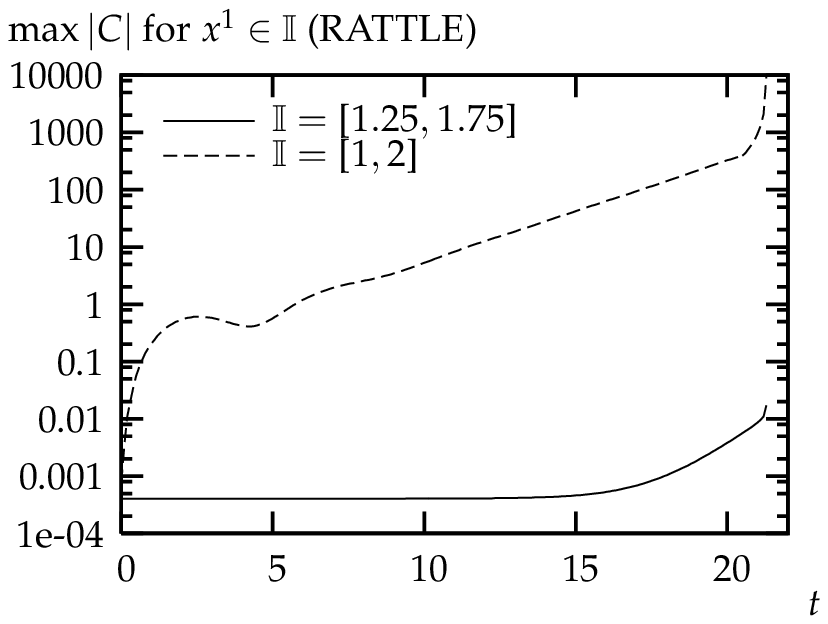}
\end{tabular}
\caption{The maximum of the Hamilton constraint function in an interval $x^1\in\mathbb I$ for Schwarzschild space-time and $N=201$ grid points. We consider an interval in the interior of the computational domain, 
$\mathbb I=[1.25,1.75]$, and the whole computational domain $\mathbb I=[1,2]$. Left: results of the St\"ormer--Verlet scheme, right: results of the RATTLE scheme.}
\label{fig:exp3:Hamilton_constraint}
\end{center}
\end{figure}

Again we investigate the Hamilton constraint. It turns out that in the St\"ormer--Verlet scheme it behaves analogous to the functions $\bfh_{11}$, $\tilde\bfh$, $\bfpi^{11}$ and $\tilde\bfpi$. That is, there are high frequency errors at the boundary that propagate into the interior and amplify (see Figure \ref{fig:Schw_Hamilton_constraint_function}). In Figure \ref{fig:exp3:Hamilton_constraint} we indeed see that the Hamilton constraint functions in the interior (in the interval $R=x^1\in[1.25,1.75]$) and near the boundary are of comparable size. Moreover we see an exponential growth of the Hamilton constraint function.

In the RATTLE scheme we observe a different behavior. There the Hamilton constraint function becomes large near the boundaries, too (see Figure \ref{fig:Schw_Hamilton_constraint_function}), but the propagation into the interior is suppressed. If we compare the Hamilton constraint function in the interior and in the whole computational domain then we see from Figure \ref{fig:exp3:Hamilton_constraint} that it stays almost constant in time in the interior until the evolution is on the brink of breaking down.

\paragraph{Discussion of results.}

We have seen that in the free evolution again high frequency errors occur, whereas they are absent in the constrained evolution scheme. 
Our simple choice of boundary conditions of Section \ref{sec:boundary_conditions} leads to boundary instabilities, and so the main source of errors are the boundaries.
In the free evolution the errors at the boundaries quickly propagate into the interior of the computational domain and amplify. In the constrained scheme, the propagation into the interior is suppressed.

\section*{Acknowledgments}
We are grateful to Sascha Husa and Bernd Br\"ugmann for their helpful advice.
This work was supported by DFG grant SFB/Transregio 7 ``Gravitational Wave Astronomy''.



\begin{thebibliography}{100}

\bibitem{Alcubierre:2003pc}
M.~Alcubierre~et al.
\newblock Toward standard testbeds for numerical relativity.
\newblock {\em Class. Quant. Grav.}, 21:589, 2004.

\bibitem{hc_andersen}
H.~C. Andersen.
\newblock Rattle: A `velocity' version of the shake algorithm for molecular
  dynamics calculations.
\newblock {\em J. Comput. Phys.}, 52:24--34, 1983.

\bibitem{Anderson_York_4556}
A.~Anderson and J.~York~Jr.
\newblock Hamiltonian time evolution for general relativity.
\newblock {\em Phys. Rev. Lett.}, 81(6):1154--1157, August 1998.

\bibitem{ADM_collection2191}
R.~Arnowitt, S.~Deser, and C.~Misner.
\newblock The dynamics of general relativity.
\newblock In L.~Witten, editor, {\em Gravitation: An Introduction to Current
  Research}, pages 227--265. Wiley, New York, U.S.A., 1962.

\bibitem{ashtekar}
A.~Ashtekar.
\newblock {\em Lectures on non-perturbative canonical gravity}.
\newblock World Scientific, Singapore, 1991.

\bibitem{baker-2006-96}
J.~G.~Baker, J.~Centrella, Dae-Il~Choi, M.~Koppitz and J.~van~Meter.
\newblock Gravitational wave extraction from an inspiraling configuration of merging black holes.
\newblock {\em Physical Review Letters}, 96:111102, 2006

\bibitem{bardeen-piran-1983}
J.~M.~Bardeen and T.~Piran.
\newblock General relativistic axisymmetric rotating systems: Coordinates and equations.
\newblock {\em Physics Reports}, 96(4):205--250, 1983.


\bibitem{Berger_Garfinkle_CQG}
B.~K. Berger, D.~Garfinkle, and E.~Strasser.
\newblock New algorithm for mixmaster dynamics.
\newblock {\em Class. Quantum Grav.}, 14:L29--L36, 1997.

\bibitem{Berger_Moncrief_PhysRevD.48.4676}
B.~K. Berger and V.~Moncrief.
\newblock Numerical investigation of cosmological singularities.
\newblock {\em Phys. Rev. D}, 48(10):4676--4687, Nov 1993.

\bibitem{blanco_costa_1995}
S.~Blanco, A.~Costa, and O.~A. Rosso.
\newblock Chaos in classical cosmology (ii).
\newblock {\em General Relativity and Gravitation}, 27:1295--1307, 1995.

\bibitem{Bonazzola_PhysRevD.70.104007}
S.~Bonazzola, E.~Gourgoulhon, P.~Grandcl\'ement, and J.~Novak.
\newblock Constrained scheme for the einstein equations based on the dirac
  gauge and spherical coordinates.
\newblock {\em Phys. Rev. D}, 70(10):104007, Nov 2004.

\bibitem{brodbeck-1999-40}
O.~Brodbeck, S.~Frittelli, P.~H\"ubner and O.~A.~Reula.
\newblock Einstein's Equations with Asymptotically Stable Constraint Propagation.
\newblock {\em Journal of Mathematical Physics}, 40:909, 1999

\bibitem{brown-2006-73}
D.~Brown.
\newblock The midpoint rule as a variational--symplectic integrator in
  hamiltonian systems.
\newblock {\em Physical Review D}, 73:024001, 2006.

\bibitem{brown-2008}
J.~D.~Brown.
\newblock Strongly Hyperbolic Extensions of the ADM Hamiltonian.
\newblock http://www.citebase.org/abstract?id=oai:arXiv.org:0803.0334, 2008

\bibitem{bruegmann-2006}
B.~Br\"ugmann, J.~A.~Gonzalez, M.~Hannam, S.~Husa, U.~Sperhake, W.~Tichy.
\newblock Calibration of Moving Puncture Simulations.
\newblock {\em Phys. Rev. D}, 77:024027, 2008.

\bibitem{calabrese_1228487}
G.~Calabrese, I.~Hinder, and S.~Husa.
\newblock Numerical stability for finite difference approximations of
  Einstein's equations.
\newblock {\em J. Comput. Phys.}, 218(2):607--634, 2006.

\bibitem{campanelli-2006-96}
M.~Campanelli, C.~O.~Lousto, P.~Marronetti and Y.~Zlochower.
\newblock Accurate Evolutions of Orbiting Black-Hole Binaries Without Excision.
\newblock {\em Physical Review Letters}, 96:111101, 2006

\bibitem{CFL}
R.~Courant, K.~Friedrichs, and H.~Lewy.
\newblock On the partial difference equations of mathematical physics.
\newblock {\em IBM J.}, 11:215--234, 1967.

\bibitem{Dirac-1958}
P.~Dirac.
\newblock The theory of gravitation in hamiltonian form.
\newblock {\em Proc. Roy. Soc. Lond.}, A 246:333, 1958.

\bibitem{Dirac-1959}
P.~Dirac.
\newblock Fixation of coordinates in the hamiltonian theory of gravitation.
\newblock {\em Phys. Rev.}, 114:924, 1959.

\bibitem{franke-2006-148}
V.~A. Franke.
\newblock {Different canonical formulations of Einstein's theory of gravity}.
\newblock {\em Theor. Math. Phys.}, 148:995--1010, 2006.

\bibitem{Fr08}
J.~Frauendiener.
\newblock {The applicability of constrained symplectic integrators in general relativity}.
\newblock {\em J. Phys. A: Math. Theor.}, 41 No 38:382005, 2008.

\bibitem{gambini-2005}
R.~Gambini and J.~Pullin.
\newblock Consistent Discrete Space-Time.
\newblock In A. Ashtekar, editor, {\em 100 YEARS OF RELATIVITY -- Space-Time Structure: Einstein and Beyond}. World Scientific, 2005.

\bibitem{gourgoulhon-2007}
E.~Gourgoulhon.
\newblock {3+1 Formalism and Bases of Numerical Relativity,
  arXiv:gr-qc/0703035}.
\newblock 2007.

\bibitem{Gundlach_2006}
C.~Gundlach, and J.~M.~Mart\'in-Garc\'ia.
\newblock {Hyperbolicity of second-order in space systems of evolution equations}.
\newblock {\em Class. Quant. Grav.}, 23: S387-S404, 2006.

\bibitem{HaLW03}
E.~Hairer, C.~Lubich, and G.~Wanner.
\newblock {Geometric numerical integration illustrated by the
St\"ormer--Verlet method.}
\newblock {\em Acta Numerica}, 12:399--450, 2003.


\bibitem{HaLW06}
E.~Hairer, C.~Lubich, and G.~Wanner.
\newblock {\em {Geometric numerical integration. Structure-preserving
  algorithms for ordinary differential equations. 2nd ed.}}
\newblock {Springer Series in Computational Mathematics 31. Berlin: Springer.},
  2006.

\bibitem{Laguna-2006}
F.~Herrmann, I.~Hinder, D.~Shoemaker and P.~Laguna.
\newblock Unequal mass binary black hole plunges and gravitational recoil.
\newblock {\em Class. Quantum Grav.}, 24:S33--S42

\bibitem{LeR04}
B.~Leimkuhler and S.~Reich.
\newblock {\em Simulating Hamiltonian Dynamics.}
\newblock {Cambridge Monographs on Applied and Computational Mathematics 14.
Cambridge: Cambridge University Press.}
2004.

\bibitem{lindblom-2006-23}
L.~Lindblom, M.~A.~Scheel, L.~E.~Kidder, R.~Owen and O.~Rinne.
\newblock A New Generalized Harmonic Evolution System.
\newblock {\em Classical and Quantum Gravity}, 23:S447, 2006

\bibitem{pretorius-2005-95}
F.~Pretorius.
\newblock Evolution of binary black hole spacetimes.
\newblock {\em Phys. Rev. Lett.}, 95:121101, 2005.

\bibitem{pretorius-2006-23}
F.~Pretorius.
\newblock Simulation of Binary Black Hole Spacetimes with a Harmonic Evolution Scheme.
\newblock {\em Class. Quant. Grav.}, 23:S529, 2006

\bibitem{Regge_Teitelboim}
T.~Regge and C.~Teitelboim.
\newblock Role of surface integrals in the hamiltonian formulation of general
  relativity.
\newblock {\em Ann. Phys.}, 88:286, 1974.

\bibitem{SSC94}
J.~M.~Sanz-Serna and M.~P.~Calvo.
\newblock {\em Numerical Hamiltonian Problems}.
\newblock Chapman \& Hall, London, 1994.

\end{thebibliography}
\end{document}